\documentclass[iop]{emulateapj} 

\usepackage {epsf}
\usepackage{epstopdf}
\usepackage {amssymb,amsmath}

\gdef\1054{MS\,1054--03}

\def\farcs{\hbox{$.\!\!^{\prime\prime}$}}
\def\simgeq{{\raise.0ex\hbox{$\mathchar"013E$}\mkern-14mu\lower1.2ex\hbox{$\mathchar"0218$}}} 

\begin {document}

\title {A CANDELS - 3D-HST synergy: resolved star formation patterns at $0.7 < z < 1.5$}

\author{Stijn Wuyts\altaffilmark{1},
Natascha M. F\"{o}rster Schreiber\altaffilmark{1}, 
Erica J. Nelson\altaffilmark{2},
Pieter G. van Dokkum\altaffilmark{2},
Gabe Brammer\altaffilmark{3},
Yu-Yen Chang\altaffilmark{4},
Sandra M. Faber\altaffilmark{5},
Henry C. Ferguson\altaffilmark{6},
Marijn Franx\altaffilmark{7},
Mattia Fumagalli\altaffilmark{7},
Reinhard Genzel\altaffilmark{1},
Norman A. Grogin\altaffilmark{6},
Dale D. Kocevski\altaffilmark{8},
Anton M. Koekemoer\altaffilmark{6},
Britt Lundgren\altaffilmark{9},
Dieter Lutz\altaffilmark{1},
Elizabeth J. McGrath\altaffilmark{10},
Ivelina Momcheva\altaffilmark{2},
David Rosario\altaffilmark{1},
Rosalind E. Skelton\altaffilmark{11},
Linda J. Tacconi\altaffilmark{1},
Arjen van der Wel\altaffilmark{4},
Katherine E. Whitaker\altaffilmark{12}}

\altaffiltext{1}{Max-Planck-Institut f\"{u}r extraterrestrische Physik, Postfach 1312, Giessenbachstr., D-85741 Garching, Germany; swuyts@mpe.mpg.de}
\altaffiltext{2}{Astronomy Department, Yale University, New Haven, CT 06511, USA}
\altaffiltext{3}{European Southern Observatory, Alonson de C\'ordova 3107, Casilla 19001, Vitacura, Santiago, Chile}
\altaffiltext{4}{Max-Planck-Institut f\"{u}r Astronomie, K\"{o}nigstuhl 17, D-69117 Heidelberg, Germany}
\altaffiltext{5}{UCO/Lick Observatory, Department of Astronomy and Astrophysics, University of California, Santa Cruz, CA 95064, USA}
\altaffiltext{6}{Space Telescope Science Institute, 3700 San Martin Drive, Baltimore, MD 21218, USA}
\altaffiltext{7}{Leiden Observatory, Leiden University, P.O. Box 9513, 2300 RA Leiden, The Netherlands}
\altaffiltext{8}{Department of Physics and Astronomy, University of Kentucky, Lexington, KY 40506, USA}
\altaffiltext{9}{Department of Astronomy, University of Wisconsin-Madison, Madison, WI 53706, USA}
\altaffiltext{10}{Department of Physics and Astronomy, Colby College, Waterville, ME 0490, USA}
\altaffiltext{11}{South African Astronomical Observatory, Observatory Road, 7925 Cape Town, South Africa}
\altaffiltext{12}{Astrophysics Science Division, Goddard Space Flight Center, Greenbelt, MD 20771, USA}

\begin{abstract}
We analyze the resolved stellar populations of 473 massive star-forming galaxies at $0.7 < z < 1.5$, with multi-wavelength broad-band imaging from CANDELS and H$\alpha$ surface brightness profiles at the same kiloparsec resolution from 3D-HST.  Together, this unique data set sheds light on how the assembled stellar mass is distributed within galaxies, and where new stars are being formed.  We find the H$\alpha$ morphologies to resemble more closely those observed in the ACS $I$ band than in the WFC3 $H$ band, especially for the larger systems.  We next derive a novel prescription for H$\alpha$ dust corrections, which accounts for extra extinction towards HII regions.  The prescription leads to consistent SFR estimates and reproduces the observed relation between the H$\alpha$/UV luminosity ratio and visual extinction, both on a pixel-by-pixel and on a galaxy-integrated level.  We find the surface density of star formation to correlate with the surface density of assembled stellar mass for spatially resolved regions within galaxies, akin to the so-called 'main sequence of star formation' established on a galaxy-integrated level.  Deviations from this relation towards lower equivalent widths are found in the inner regions of galaxies.  Clumps and spiral features, on the other hand, are associated with enhanced H$\alpha$ equivalent widths, bluer colors, and higher specific star formation rates compared to the underlying disk.  Their H$\alpha$/UV luminosity ratio is lower than that of the underlying disk, suggesting the ACS clump selection preferentially picks up those regions of elevated star formation activity that are the least obscured by dust.  Our analysis emphasizes that monochromatic studies of galaxy structure can be severely limited by mass-to-light ratio variations due to dust and spatially inhomogeneous star formation histories.
\end{abstract}


\section {Introduction}
\label{intro.sec}

Studies of galaxy evolution from the peak of cosmic star formation to the present day have matured tremendously over the past two decades.  Initially, the efforts focussed on the detection and selection of distant blobs of light by means of color selections (Steidel et al. 1996; Franx et al. 2003; Daddi et al. 2004), of which generally brighter subsets were confirmed spectroscopically.  Gradually, galaxy-integrated photometry over ever increasing wavelength baselines, from the rest-frame UV to infrared, allowed inferences on the global stellar population properties such as stellar mass, age and star formation rate (SFR; Papovich et al. 2001; F\"{o}rster Schreiber et al. 2004; Shapley et al. 2005).  Alongside developments on the stellar population front, deep field observations carried out with the Hubble Space Telescope (HST) have played a pivotal role in opening up the distant universe for resolved, morphological analyses (e.g., Williams et al. 1996; Giavalisco et al. 2004; Rix et al. 2004; Koekemoer et al. 2007).  Initially, such lookback surveys were predominantly monochromatic, or at most spanning a limited number of bands probing the rest-frame UV emission, which is dominated by young O and B stars.

The first HST-based extensions into the rest-frame optical of high-redshift galaxies already date back more than a decade (Thompson et al. 1999; Dickinson et al. 2000), but until recently were limited to small sample sizes due to the small field of view of the NICMOS camera.  The installment of the Wide Field Camera 3 (WFC3) during Servicing Mission 4  changed the situation dramatically.  Benefitting from WFC3's enhanced sensitivity and larger field of view, the Cosmic Assembly Near-infrared Deep Extragalactic Legacy Survey (CANDELS; Grogin et al. 2011; Koekemoer et al. 2011) has been mapping the rest-optical properties of galaxies since the peak of cosmic star formation at high resolution in 5 disjoint fields on the sky (GOODS-South, GOODS-North, COSMOS, UDS, EGS).  Exploiting initial subsets of the CANDELS legacy data set, several works have contrasted monochromatic structural measurements to galaxy-integrated stellar population properties (Wuyts et al. 2011b; Szomoru et al. 2011; Bell et al. 2012; Wang et al. 2012; Kartaltepe et al. 2012; Bruce et al. 2012).  Overall, these authors found strong correlations between structure and stellar populations (i.e., a 'Hubble sequence') to be present out to at least $z \sim 2.5$ (see also Franx et al. 2008; Kriek et al. 2009a).  Star-forming galaxies residing on the star formation rate (SFR) - mass  'main sequence' relation (Noeske et al. 2007; Elbaz et al. 2007, 2011; Daddi et al. 2007) tend to be the largest at their mass, and are best characterized by exponential disk profiles, while quiescent galaxies lying below the main sequence have higher Sersic indices, i.e., are more bulge-dominated (Wuyts et al. 2011b; although see Barro et al. 2013 for interesting sub-populations deviating from this overall trend).  This implies bulge growth and quenching are intimately connected.

The molecular gas content of normal star-forming galaxies (SFGs) is known to increase rapidly with lookback time, to gas mass fractions of $\sim 0.33$ (0.47) at $z \sim 1.2$ (2.2) (Tacconi et al. 2010, 2013; Daddi et al. 2010).  Simple stability arguments predict such gas-rich systems to be prone to gravitational collapse on scales of $\sim 1$ kpc.  Indeed, such features, and more generally irregular morphologies, have been abundantly reported in high-redshift SFGs (albeit on monochromatic, and often rest-UV or H$\alpha$ images; see, e.g., Elmegreen et al. 2004, 2009a; Genzel et al. 2008, 2011).  Analytic work as well as hydrodynamic simulations have proposed that the giant clumps may migrate inward via dynamical friction and tidal torques, thereby providing an alternative channel to bulge formation to the conventional merger scenario (e.g., Noguchi 1999; Immeli et al. 2004a,b; Bournaud et al. 2007; Dekel et al. 2009).  In part, the efficiency of such violent disk instabilities in driving significant bulge growth is subject to the longevity of the clumps, which may be limited by vigorous outflows driven by internal star formation (e.g., Newman et al. 2012; Genel et al. 2012; Hopkins et al. 2012).  However, several studies argued that even if the clumps do not remain bound, torques in unstable disks will still lead to an enhanced gas inflow rate with respect to stable configurations (Krumholz \& Burkert 2010; Bournaud et al. 2011; Genel et al. 2012; Cacciato et al. 2012).

Additional clues on the emergence of bulges and role or nature of clumps can come from studies of the resolved stellar populations within galaxies.  Here too, ACS+WFC3 imaging from CANDELS and pre-existing campaigns offer the capability to expand sensitively on pioneering work by, e.g., Abraham et al. (1999), Elmegreen et al. (2009b) and F\"{o}rster Schreiber et al. (2011a,b).  Wuyts et al. (2012) carried out 7-band stellar population modeling on a (binned) pixel-by-pixel basis for 649 massive ($\log M_* > 10$) star-forming galaxies (SFGs) at $0.5 < z < 2.5$, the largest and only mass-complete sample of SFGs subject to such a detailed analysis to date.  Translating the internal variations in intensity and color to spatial distributions of more physically relevant quantities such as stellar mass, star formation rate (SFR), age and extinction, these authors found high-redshift SFGs to be smoother and more compact in mass than in light, with color variations driven by a combination of radial extinction gradients and spatial (short-term) fluctuations in the star formation history (see also F\"{o}rster Schreiber et al. 2011a,b; Guo et al. 2012; Lanyon-Foster et al. 2012; Szomoru et al. 2013).  In particular, Wuyts et al. (2012) and Guo et al. (2012) found regions with enhanced surface brightness with respect to the underlying disk to be characterized by enhanced levels of star formation and younger ages than interclump regions at the same galactocentric distance.  Typical inferred clump ages of 100 - 200 Myr at $z \sim 2$ imply that the clumps correspond to short-lived star-forming phenomena, possibly limited in lifetime by stellar feedback.  If inward clump migration is taking place, this should happen efficiently on timescales similar to the orbital timescale.  Radial age gradients of clumps (F\"{o}rster Schreiber et al. 2011b; Guo et al. 2012) may signal such migrational processes to be at play.  At $z \sim 1$, which is the epoch we focus on in this paper, the physics leading to regions of excess surface brightness and locally enhanced star formation may be a mix of the above violent disk instabilities frequently studied at $z \sim 2$, and more conventional processes known from nearby star-forming galaxies.  That is, gas fractions can still be sufficiently high to cause gravitational collapse, while timescales become long enough for other instabilities such as spiral density waves to arise (the latter are only in rare cases seen at $z \sim 2$, see Law et al. 2012).  We note, as we did in our previous work, that while for simplicity we occasionally use the same short-hand 'clump' terminology as Wuyts et al. (2012), the specific diagnostic identifying enhanced surface brightness regions does not discriminate between round knots or spiral features.

Despite the richness of the HST broad-band data sets, the multi-wavelength sampling available at high resolution is still modest in comparison to state-of-the-art spectral energy distributions (SEDs) on a galaxy-integrated level.  The latter span up to 30+ medium and broad bands from 0.3 to 8 micron (e.g., Ilbert et al. 2009; Whitaker et al. 2011; Spitler et al. 2012), complemented further by mid-to-far-infrared photometry, probing dust-obscured star formation within at least the more massive SFGs (e.g., Magnelli et al. 2013).  The performance and limitations of broad-band SED modeling as a tool to infer stellar population properties have been abundantly tested on synthetic observations of simulated galaxies (Wuyts et al. 2009; Lee et al. 2009; Pforr et al. 2012; Mitchell et al. 2013).  It is clear that, while 4 to 7 HST broad bands may be sufficient to reconstruct stellar mass distributions\footnote{Stellar mass estimates to an accuracy of $\sim 0.2$ dex can arguably even be obtained on the basis of two broad bands, provided they bracket the Balmer/4000\AA\ break (Bell \& de Jong 2001; Taylor et al. 2011; and see F\"{o}rster Schreiber et al. 2011a for an application of this technique to the resolved mass distribution in a $z \sim 2$ galaxy).}, some level of degeneracy (e.g., between age and dust extinction) will be inherent to the inferences made on a pixel-by-pixel basis when SEDs span the relatively narrow wavelength range from observed $B$ to $H$ band.  Additional empirical constraints would allow us to test and build on the more model-sensitive findings from Wuyts et al. (2012).

The 3D-HST legacy program (van Dokkum et al. 2011; Brammer et al. 2012) and GOODS-North grism survey GO-11600 (PI B. Weiner), together covering all five CANDELS fields, provide two such empirical constraints.  First, the grism data yields spectroscopic redshifts for thousands of galaxies, eliminating an important source of uncertainty that propagates in all derived stellar population properties.  Second, each dispersed galaxy image can be considered as a continuum with superposed resolved line emission maps at HST resolution.  After subtracting the continuum underlying the H$\alpha$ emission of $z \sim 1$ galaxies, we are therefore left with an independent SFR diagnostic probing the same kiloparsec scales as the CANDELS multi-wavelength broad-band imaging.  The additional information furthermore allows us to better constrain the effects of extinction.  Nelson et al. (2012; 2013) demonstrated the power of this technique by contrasting the stacked one-dimensional H$\alpha$ and $H$-band surface brightness profiles.  They found the extent of the ionized gas and stellar light profiles to be similar for small galaxies.  The ionized gas component of the larger galaxies on the other hand, exhibit larger scale lengths than the stellar component, consistent with an inside-out growth scenario.  As is the case for rest-optical light profiles, the H$\alpha$ emission of star-forming galaxies in their sample is best characterized by an exponential disk profile.

In this paper, we build on the analyses by Nelson et al. (2012; 2013) and Wuyts et al. (2012) by combining the resolved H$\alpha$ information on $z \sim 1$ SFGs with pixel-by-pixel stellar population modeling of their multi-wavelength broad-band photometry.  The present work has strong connections to a very comprehensive exploration of H$\alpha$ and SFR profiles throughout the two-dimensional SFR-Mass space (E. Nelson et al. in prep).  While restricting the analysis to systems with well-detected line emission, we expand the sample with respect to previous work by exploiting the data in all five CANDELS/3D-HST fields (see the sample description in Section\ \ref{obs_sample.sec}).  We discuss the methodology to derive stellar population properties such as star formation and surface mass density in Section\ \ref{methodology.sec}.  Here, we place a special emphasis on dust corrections, and in particular extra extinction towards the HII regions from which the H$\alpha$ line emission originates, using UV+IR based SFRs as an additional calibrator.  After a visual impression of characteristic features in the H$\alpha$ and broad-band images (Section\ \ref{examples.sec}), we quantify the correspondence between H$\alpha$ and broad-band morphologies at different wavelengths (Section\ \ref{crosscor.sec}), compare emission line diagnostics to stellar population properties derived from broad-band information alone (Section\ \ref{broad_vs_line.sec}), and finally contrast the $H\alpha$ and corresponding star formation properties of galaxy centers, clumps/spiral arms, and underlying disks within our sample of star-forming galaxies with $\log M_* > 10$ (Section\ \ref{prof2D.sec}).

Throughout this paper, we quote magnitudes in the AB system, assume a Chabrier (2003) initial mass function (IMF) and
adopt the following cosmological parameters: $(\Omega _M, \Omega_{\Lambda}, h) = (0.3, 0.7, 0.7)$.

\section{Observations and Sample}
\label{obs_sample.sec}

HST broad-band imaging from CANDELS and pre-existing surveys, together with HST grism spectroscopy from 3D-HST form the core data set on which this paper is based.  In addition, our sample definition makes use of the wealth of ancillary data collected in the five CANDELS/3D-HST fields, including multiple medium- and broad-band imaging campaigns from the ground, and space-based photometry from Spitzer/IRAC, Spitzer/MIPS, and Herschel/PACS.  We derived the galaxy-integrated stellar masses and SFRs based thereupon following identical procedures to Wuyts et al. (2011b).  That is, the SFRs that enter the sample selection criteria are based on a ladder of SFR indicators, using UV+PACS for PACS-detected galaxies, otherwise UV+MIPS 24$\mu$m for MIPS-detected galaxies, and finally SFRs from SED modeling (that also yields the stellar masses) for sources without IR detection.  The 3D-HST catalogs with consistent multi-wavelength photometry that serve as input to the galaxy-integrated SED modeling are presented by Skelton et al. (2013).

\subsection{CANDELS HST imaging}
\label{CANDELS.sec}

CANDELS provides deep WFC3 imaging of five disjoint fields on the sky: GOODS-South, GOODS-North, COSMOS, UDS, and EGS, totaling approximately 800 arcmin$^2$.  Point-source limiting depths vary from H = 27.0 mag in CANDELS/Wide to H = 27.7 mag in CANDELS/Deep (the central halves of the GOODS fields).  We refer the reader to Grogin et al. (2011) for an overview of the survey layout and Koekemoer et al. (2011) for details on the data reduction.  Other HST imaging that enters our analysis includes the GOODS ACS campaigns (Giavalisco et al. 2004) and subsequent ACS epochs taken as part of the HST supernovae search, COSMOS/ACS (Koekemoer et al. 2007), AEGIS/ACS (Davis et al. 2007), and ERS/WFC3 (PI O'Connell) observations.  Together, these yield $B_{435}V_{606}i_{775}z_{850}J_{125}H_{160}$ photometry in the GOODS fields, with in addition $Y_{098}$ coverage of the ERS region (top third of GOODS-South), and $Y_{105}$ coverage of the CANDELS/Deep regions.  For the remaining fields, spectral energy distributions are sampled by $V_{606}I_{814}J_{125}H_{160}$ photometry.

\subsection{3D-HST grism spectroscopy}
\label{3DHST.sec}

Brammer et al. (2012) describe in detail the specifications of the 3D-HST legacy program.  Briefly, 3D-HST\footnote{The G141 grism coverage of the GOODS-North field from program GO-11600 (PI: B. Weiner) is incorporated into 3D-HST as the observational strategy is nearly identical to that of 3D-HST.} covers three quarters (625 arcmin$^2$) of the CANDELS treasury survey area, and naturally our sample will be drawn from the region where the two data sets overlap.  In this paper, we make use of the two orbit depth WFC3/G141 grism exposures, but note that in parallel two to four orbits of ACS/G800L grism data were taken as part of the program.  The WFC3/G141 grism data reach a 5$\sigma$ emission-line sensitivity of $\sim 5 \times 10^{-17}\ erg\ s^{-1}\ cm^{-2}$, and cover a wavelength range from 1.1 to 1.6 $\mu$m.  We make use of the v2.1 internal release by the 3D-HST team, which differs most significantly from the data handling described by Brammer et al. (2012) in that the half-pixel dithered exposures are combined by interlacing rather than drizzling, thereby reducing the correlated noise.  The extraction of grism spectra followed the steps outlined by Brammer et al. (2012).  As part of the extraction pipeline, a spectral model is constructed that, once convolved with the F140W pre-image of the galaxy that serves as a template, best fits the observed grism spectrum.  In this process, neighbors are treated simultaneously to reduce contamination effects.  Redshifts are fitted to the combination of available broad-band and grism information.  In practice, for the sample of line-emitting galaxies analyzed in this paper, the ancillary broad-band data only serves in the redshift determination to prevent line misidentifications.

\subsection{Sample definition} 
\label{sample.sec}

\begin {figure}[t]
\plotone{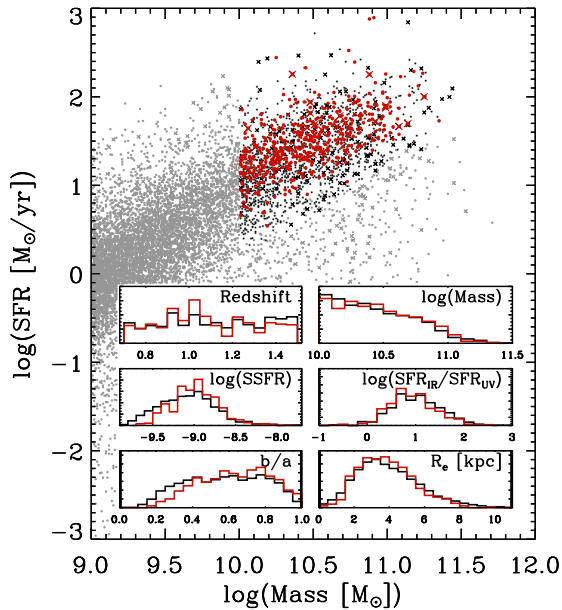}  
\caption{
Location of our massive SFG sample ({\it red}) in SFR-Mass space, overplotted on the distribution of the overall galaxy population at $0.7 < z < 1.5$.  Crosses mark X-ray detected sources.  Inset panels show the distribution in redshift, stellar mass ([$M_\sun$]), specific star formation rate ([yr$^{-1}$]), and obscuration level ($SFR_{IR}/SFR_{UV}$) for the H$\alpha$ sample analyzed in this paper ({\it red}) compared to the complete sample of star-forming galaxies above $\log M_* = 10$ (black).
\label{sample.fig}}
\end {figure} 

\begin {deluxetable*}{lrrrrrr}
\tablecolumns{7}
\tablewidth{\linewidth}
\tablecaption{Sample selection of massive SFGs at $0.7 < z < 1.5$ with H$\alpha$ maps from 3D-HST}
\tablehead{
\colhead{Selection} & \colhead{N(total)} & \colhead{N(GOODS-S)} & \colhead{N(GOODS-N)}
 & \colhead{N(EGS)} & \colhead{N(UDS)} & \colhead{N(COSMOS)}
} 
 \startdata
 $0.7 < z < 1.5$ \& $\log M_* > 10$       & 3115 & 609 & 656 & 340 & 692 & 818 \\
 ... \& $SSFR > 1/t_{\rm Hubble}$       & 1844 & 359 & 381 & 206 & 433 & 465 \\
 ... \& H$\alpha$ $S/N > 8$      & 636 & 109 & 176 & 96 & 130 & 125 \\
 ... \& no contamination or continuum residuals   & 506 & 95 & 143 & 72 & 110 & 86 \\
 ... \& Final sample\tablenotemark{a}   & 473 & 85 & 129 & 69 & 108 & 82 \\
\enddata
 
\tablenotetext{a}{\footnotesize \looseness=-2 Compact or heavily point source dominated sources with a X-ray counterpart are excluded from the final sample.}
  \label{sample.tab} 
\end {deluxetable*}

The adopted redshift range of  $0.7 < z < 1.5$ for our sample is dictated by the requirement that the H$\alpha$ emission falls within the WFC3/G141 wavelength coverage.  We next apply the same basic criteria to select massive SFGs as Wuyts et al. (2012), namely $\log M_* > 10$ and a specific SFR $SSFR > 1 / t_{\rm Hubble}$.  The depth of the $H$-band selected parent catalogs guarantees completeness down to this mass limit (and in fact more than an order of magnitude below).  Over the entire CANDELS/3D-HST area with coverage by the G141 grism spectroscopy and at least 4 HST broad bands, this amounts to 1844 massive $z \sim 1$ SFGs.

Subsequent selection criteria serve to optimize the quality of our inferences on resolved H$\alpha$ properties, at the expense of compromising the mass completeness of our SFG sample.  For our final sample, we require galaxies to have secure redshifts, with well-covered H$\alpha$ emission detected at the $8\sigma$ significance level.  We furthermore apply a conservative screening against objects whose grism spectra are contaminated by neighboring sources, as identified based on products of the 3D-HST pipeline, supplemented with a visual inspection.  In this step, we also weed out a small number of sources with prominent residuals over the full wavelength range after subtraction of the continuum model produced by the 3D-HST extraction and fitting pipeline.  Finally, we exclude 33 bright X-ray detected objects for which the compact nature of their HST imaging could hint at a significant non-stellar contribution to both the broad-band and the H$\alpha$ line emission.  Before the latter cut, the overall fraction of X-ray detected sources in the H$\alpha$ sample exceeds the respective X-ray detected fraction determined for the complete parent sample of massive SFGs by a factor 1.5 ($\sim 12\%$ versus $\sim 8\%$\footnote{The percentages quoted are for the sum of all 5 CANDELS/3D-HST fields.  While the absolute percentages are field dependent due to varying depth of the X-ray imaging, the relative factor 1.5 between the H$\alpha$ and parent sample is similar for all fields.}).  A possible explanation for this difference may be a tendency for actively star-forming, line-emitting galaxies to more frequently host AGN activity (e.g., Santini et al. 2012; Trump et al. 2013).  In addition, since the H$\alpha$ and $[NII]$ emission lines are blended in the WFC3 grism data (see Section\ \ref{Halpha_extraction.sec}), the enhanced $[NII]/H\alpha$ ratios of AGN-hosting galaxies may push more of them over the signal-to-noise threshold for line emission.  In our final sample, the only X-ray detected sources remaining are located in GOODS-South (16) and GOODS-North (7).  We note that in these deep X-ray fields (4 Ms and 2 Ms exposure, respectively), not all X-ray detected sources are necessarily AGN.  Moreover, no evidence for point source contamination is seen in their HST imaging.  We break down our final sample and the intermediate selection steps by field in Table\ \ref{sample.tab}.

It is important to note that our final sample of 473 massive SFGs spans the entire range in mass and SFR of typical main-sequence galaxies above $\log(M_*) = 10$ (see Figure\ \ref{sample.fig}).  The inset panels in Figure\ \ref{sample.fig} present a closer look at the relation between the H$\alpha$ sample and the underlying complete parent population of massive SFGs at $0.7 < z < 1.5$.  To first order, the samples are well matched in redshift, mass, star formation activity, obscuration ($SFR_{IR}/SFR_{UV}$), axial ratio and size, in terms of dynamic range spanned as well as the distribution within that range.  In more detail, the H$\alpha$ sample shows subtle biases against the lowest SSFR systems, against the most obscured galaxies at a given SFR, and, related, against the most edge-on galaxies.  The similarity between the two samples is encouraging, as it implies that the analysis of the H$\alpha$ sample presented in this paper can reveal generic insights for the entire massive SFG population during the transition from the peak of cosmic star formation to the initial phases of its decline.

\section {Methodology}
\label{methodology.sec}

\subsection{Modeling of the broad-band SEDs}
\label{SEDmodeling.sec}

The three basic steps towards resolved SED modeling are PSF matching, pixel binning, and stellar population synthesis modeling of the individual spatial bins.  Each of these steps is described in depth by Wuyts et al. (2012).  Briefly, we work at the WFC3 $H_{160}$ resolution of $0 \farcs 18$, and used the Iraf PSFMATCH algorithm to build kernels to bring all shorter wavelength images to the same PSF width and shape.  We next applied the Voronoi binning scheme by Cappellari \& Coppin (2003) to group neighboring pixels together in bins so as to achieve a minimum signal-to-noise level of 10 per bin.  We applied the Voronoi binning scheme to the WFC3 $H_{160}$ maps.  Photometry in the other (PSF-matched) ACS and WFC3 bands was measured within the identical bins of grouped pixels in order to guarantee that the colors entering our analysis probe consistently the same physical regions.  The resulting multi-wavelength photometry per spatial bin is then fed to EAZY (Brammer et al. 2008) to derive rest-frame photometry and FAST (Kriek et al. 2009b) to fit stellar population synthesis models from Bruzual \& Charlot (2003), with identical settings to Wuyts et al. (2011b; 2012).  Since all galaxies in our sample have spectroscopically confirmed redshifts from the H$\alpha$ line detection in the grism data (and in half of the cases confirmed independently by ground-based spectroscopic campaigns), we fix the redshift to its spectroscopically determined value in our SED fitting, hence reducing the number of free parameters by one.  In addition, we assume the stars have a fixed, solar metallicity.  While this assumption is frequently adopted in the literature on stellar populations of massive $z \sim 1$ galaxies, we caution that additional constraints and further investigations are needed to address the validity and impact of this assumption for resolved studies on kiloparsec scales.  We follow Salim et al. (2007) in adopting parameter values marginalized over the multi-dimensional grid explored in SED fitting, as this approach proved most robust (compared to, e.g., adopting the least squares solution) in the limit of poorly sampled SEDs.  Additional notes on the reliability of the broad-band SED modeling, and its impact on the results presented in this paper are discussed in the Appendix.

\subsection{Extracting H$\alpha$ emission line maps}
\label{Halpha_extraction.sec}

In order to extract H$\alpha$ emission line maps, we follow Nelson et al. (2012; 2013), and subtract the continuum model from the observed 2D grism spectrum.  The resulting residual image is then mapped to the CANDELS frame using redshift and astrometric information, and contains the emission line surface brightness distribution without imposed prior on its morphology.  At the spectral resolution of $\delta v \approx 1000$ km s$^{-1}$, H$\alpha$ and [NII] $\lambda\lambda$6548+6583 are blended.  As we lack additional constraints, we apply throughout the paper a simple downscaling of the observed emission line flux by a factor 1.2, and refer to this quantity as the H$\alpha$ flux.  In reality, [NII]/H$\alpha$ ratios may vary between galaxies as well as spatially within galaxies (see, e.g., Liu et al. 2008; Yuan et al. 2012, 2013; Queyrel et al. 2012; Swinbank et al. 2012; Jones et al. 2013; F\"orster Schreiber et al. 2013).  However, restricting the above spectroscopic samples to the same redshift range as adopted in this paper, the scatter in [NII]/H$\alpha$ is substantial compared to any systematic trend, if present, with galaxy mass above $\log(M_*)=10$.  Furthermore, the [NII]/H$\alpha$ gradients reported based on adaptive optics assisted observations are typically shallow (F\"orster Schreiber et al. 2013).  We conclude that a higher order correction than the uniform scaling factor we apply is not justified by present data. 

Again following Nelson et al. (2013), we apply a wedge-shaped mask to regions that could potentially be affected by [SII] $\lambda\lambda$6716+6731 line emission (redward from H$\alpha$ along the dispersion axis), which can mimic the appearance of an off-center clump in the H$\alpha$ maps.

\subsection{Dust corrections to the H$\alpha$ emission}
\label{Halpha_dust.sec}

\subsubsection{Birth clouds and diffuse interstellar dust}

Proper extinction corrections are critical for the physical interpretation of dust-sensitive SFR tracers.  This applies to short wavelength (rest-UV) broad-band indicators, motivating to a large extent programs such as CANDELS.  While at the longer, rest-optical wavelengths attenuation laws predict a significantly suppressed impact by dust, this may not be the case for the H$\alpha$ emission at 6563\AA, depending on the geometry of dust and stars.  H$\alpha$ emission originates from HII regions immediately surrounding young star-forming regions, that are known to be often associated with enhanced levels of obscuring material.  As such, the nebular emission emerging near massive O stars that have not yet dispersed or escaped from their dust-rich birth clouds, will be subject to extra extinction with respect to continuum light at the same wavelength that is produced by the bulk of the stars (no longer embedded in the molecular clouds in which they once formed).  The observed (i.e., attenuated) H$\alpha$ flux then relates to the intrinsic flux $F_{H\alpha,\ int}$ as

\begin {equation}
F_{H\alpha,\ obs} = F_{H\alpha,\ int}\ 10^{-0.4 A_{cont}}\ 10^{-0.4 A_{extra}}
\end {equation}

where $A_{cont}$ represents the attenuation by diffuse dust in the galaxy and $A_{extra}$ represents the attenuation happening locally in the birth cloud.  The latter have a negligible covering factor and therefore affect no other galaxy light than that emerging from the respective star-forming region itself.  The geometrical picture sketched here relates intimately to the framework of the Charlot \& Fall (2000) model, and subsequent refinements by, e.g., Wild et al. (2011), Pacifici et al. (2012), and Chevallard et al. (2013).

Empirical evidence for the need of an extra extinction correction to H$\alpha$ (i.e., $A_{extra} \not= 0$) was first presented for a sample of nearby starburst galaxies by Calzetti et al. (1994; 2000).  Also at larger lookback times, evidence for differential extinction between nebular regions and the bulk of the stars has been mounting, from arguments based on the observed H$\alpha$ equivalent widths (van Dokkum et al. 2004; F\"orster Schreiber et al. 2009), comparisons of multi-wavelength SFR indicators (Wuyts et al. 2011a; Mancini et al. 2011), and measurements of the Balmer decrement (Ly et al. 2012; Price et al. 2013).  While showing a median consistency with the local calibration by Calzetti et al. (2000; $A_{extra} = 1.27 A_{cont}$), the uncertainties and sample sizes used in those studies did not allow the authors to discriminate between a constant $A_{extra}$ (i.e., constant birth cloud properties independent of the optical depth from the diffuse component) and one that scales proportionally with the diffuse column as $A_{extra} \propto A_{cont}$.  Conceptually, one can think of the diffuse attenuation as being determined by the galaxy's gas fraction, dust-to-gas ratio, large-scale geometry and orientation.  The attenuation from birth clouds is expected to share some of these dependencies (e.g., dust-to-gas ratio), but not others (e.g., orientation, as demonstrated by Wild et al. 2011).  If the gas fraction of a galaxy increases, all other properties remaining the same, nothing will change to $A_{extra}$ if this results merely in an increased number of star-forming clouds, each with identical conditions.  If the mass of individual clouds does increase with galaxy-integrated gas fraction, it depends on the cloud scaling relations whether this impacts $A_{extra}$ (e.g., for a Larson 1981 scaling relation $\rho_{cloud} \sim r^{-1}$ the optical depth would remain unaffected).

\subsubsection{Calibrating H$\alpha$ dust corrections using multi-wavelength SFR indicators}
\label{calibrating.sec}

\begin {figure} [htbp]
\plotone{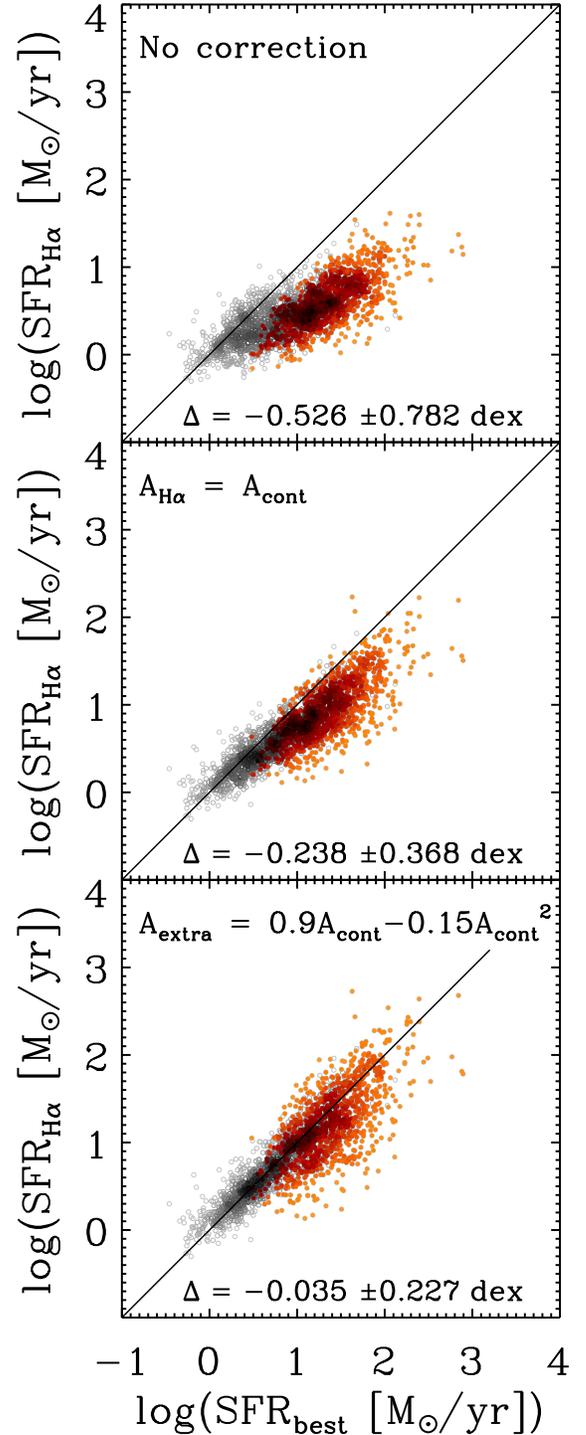}  
\caption{
Comparison of H$\alpha$-based SFR estimates to the reference ladder of SFR indicators from Wuyts et al. (2011b), in order of priority based on UV + {\it Herschel}/PACS, otherwise {\it Spitzer}/MIPS 24$\mu$m, otherwise $U$-to-8$\mu$m SED modeling.  We mark UV+IR-based SFRs in red, and indicate galaxies without IR detections with gray circles.  Crowded regions of the diagram are displayed with a darker hue.  H$\alpha$ SFRs need to be corrected for dust extinction to avoid underestimates (top panel).  Applying an extinction correction corresponding to what is inferred from SED modeling is insufficient (middle panel).  Accounting for extra extinction towards HII regions, we find a good correspondence to the reference SFRs, with modest scatter and without systematic offsets (bottom panel).
\label{SFRcomparison.fig}}
\end {figure}

\begin {figure*} [t]
\includegraphics[width=0.48\textwidth]{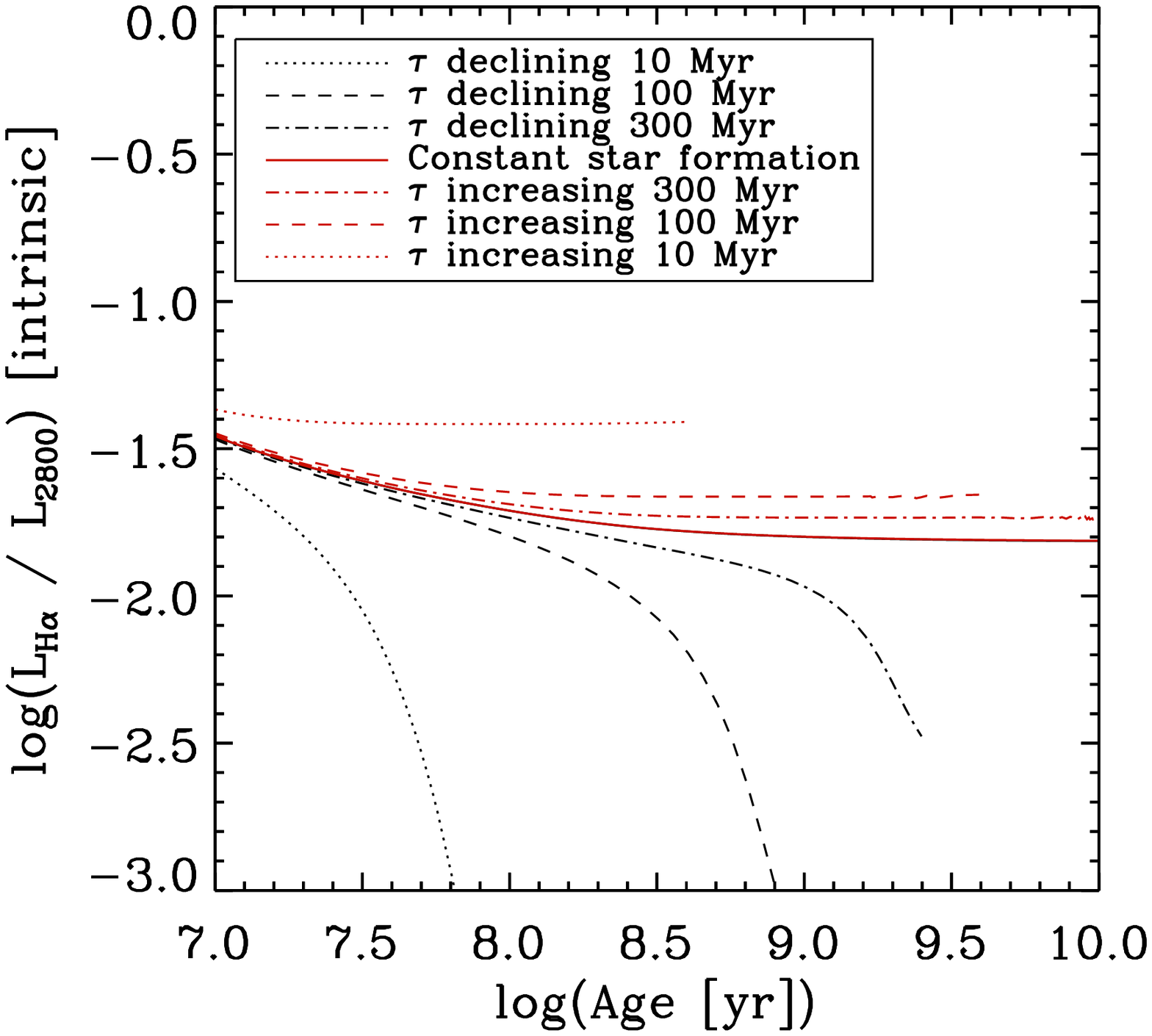}   
\includegraphics[width=0.48\textwidth]{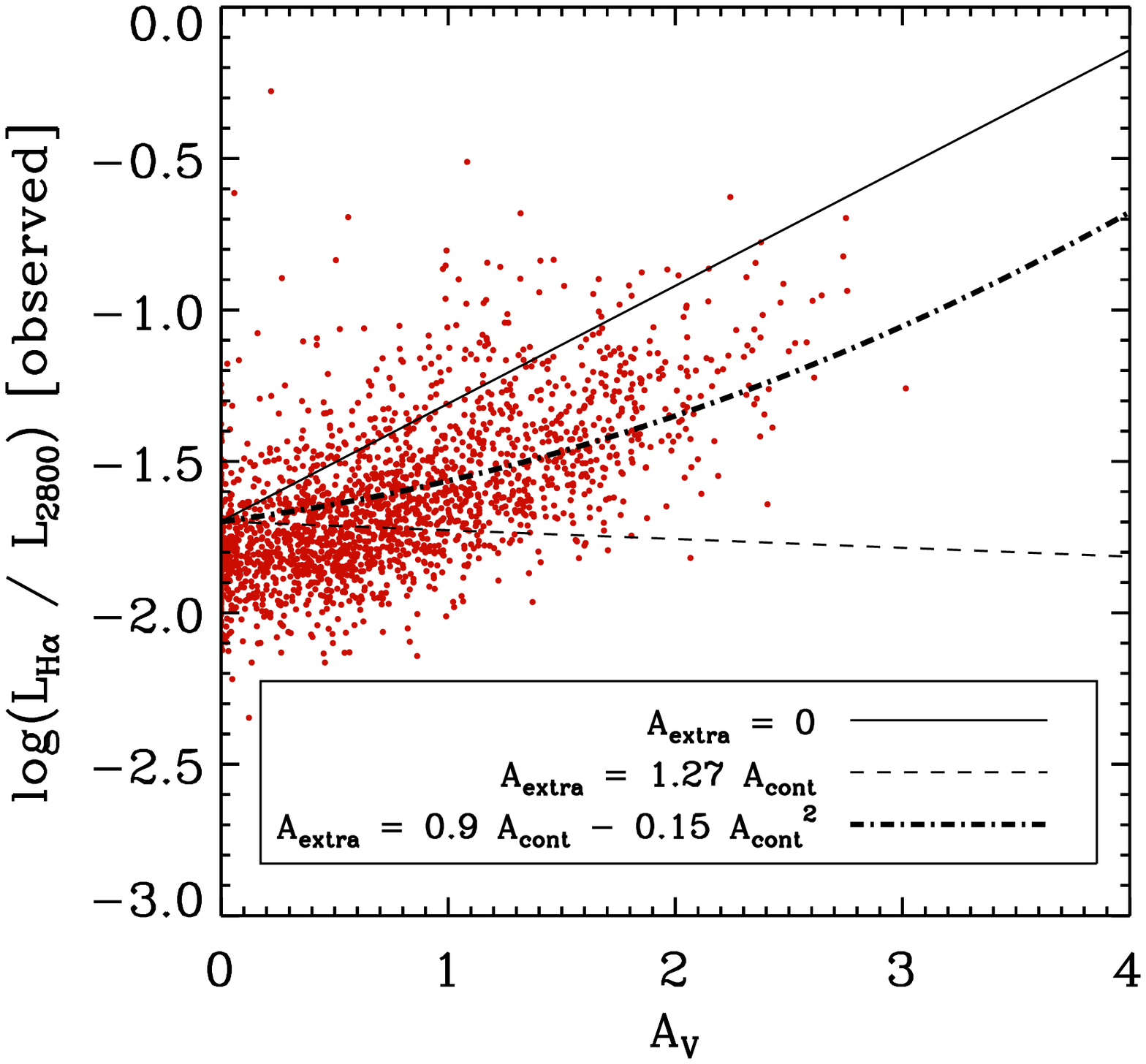}   
\caption{
{\it Left:} Dependence of the intrinsic (i.e., extinction free) H$\alpha$/UV (2800\AA) luminosity ratio on age since the onset of star formation for a range of star formation histories (exponentially declining or increasing, and constant star formation).  The Kennicutt (1998) SFR conversions from H$\alpha$ and UV assume a population with constant star formation ongoing for 100 Myr.   {\it Right:} Given a stellar population as assumed by Kennicutt (1998), the observed H$\alpha$/UV luminosity ratio depends on the visual extinction $A_V$ as indicated by the black lines, which correspond to three different models for extra extinction towards HII regions: no extra extinction ({\it solid}), double Calzetti ($A_{H\alpha} = A_{cont} / 0.44$; {\it dashed}), and limited extra extinction (Equation\ \ref{bestfit_dustcor.eq}; {\it dash-dotted}).  The latter model best describes the locus of observed galaxies ({\it small red dots}).
\label{HaUV_int.fig}}
\end {figure*}

From these considerations, we conclude that the precise functional form of $A_{extra}$ may not be a constant, or proportional to $A_{cont}$, but rather a hybrid form between those two extremes (i.e., $A_{extra}$ saturating with increasing $A_{cont}$).  By lack of other direct constraints at $z \sim 1$, we take an empirical approach to capture this behavior, and define a prescription that yields a small scatter and negligible systematic offset between two sets of galaxy-integrated SFR measurements.  As reference, we adopt the ladder of SFR indicators from Wuyts et al. (2011b), which is based on UV+IR emission, or SED modeling in case no IR detection is available (see Section\ \ref{obs_sample.sec}).   Figure\ \ref{SFRcomparison.fig} contrasts H$\alpha$-based SFRs from 3D-HST to the reference SFRs.  The top panel iterates the obvious necessity for dust corrections to the H$\alpha$ emission.  The large sample statistics exploited here also demonstrate unambiguously the presence of additional extinction towards the nebular regions (middle panel of Figure\ \ref{SFRcomparison.fig}, where $A_{extra}=0$ is assumed and significant systematic underestimates for highly star-forming systems are evident).  Finally, the bottom panel of Figure\ \ref{SFRcomparison.fig} shows the improved agreement between the SFR measurements once differential extinction is accounted for.  The adopted prescription

\begin{equation}
A_{extra} = 0.9 A_{cont} - 0.15 A_{cont}^2,
\label{bestfit_dustcor.eq}
\end {equation}

yields a scatter of 0.227 dex and a negligible systematic offset of -0.035 dex.  We note that applying the Calzetti et al. (2000) prescription for extra extinction (corresponding to $A_{H\alpha} = A_{cont} / 0.44$, or equivalently $A_{extra} = 1.27 A_{cont}$) leads to a slightly higher systematic offset of 0.066 dex, and produces a larger scatter of 0.284 dex.  If we were to adopt the Calzetti et al. (2000) prescription, all objects with high inferred $A_V$ values from broad-band SED modeling would have dust-corrected H$\alpha$ SFRs systematically in excess of the reference indicator.  We note that the need for an extra extinction correction is a result that is largely driven by the more actively star-forming galaxies.  For most of them, UV+IR (i.e., bolometric, rather than dust corrected) measurements of the SFR are available.  Those galaxies that lack IR detections in the CANDELS/3D-HST fields generally suffer less obscuration and are therefore less sensitive to the precise dust correction applied (although applying a constant $A_{extra}$, even to sources with $A_{cont}$ near zero, would lead to overestimated H$\alpha$ SFRs at the low-SFR end).

\subsubsection{H$\alpha$/UV ratios}

We now assess the validity of the calibration to correct H$\alpha$ for dust extinction (Equation\ \ref{bestfit_dustcor.eq}) by considering an independent dust-sensitive diagnostic, namely the H$\alpha$/UV luminosity ratio.  Here, we define the UV luminosity as the rest-frame luminosity $L_{2800} \equiv \nu L_\nu(2800\AA)$, which we compute with EAZY.  In principle, the $L_{H\alpha}/L_{UV}$ ratio is not uniquely dependent on dust extinction, but may also vary among galaxies that differ in their relative number of O and B stars, due to differences in the star formation history or initial stellar mass function (see, e.g., Meurer et al. 2009).  The reason is that the H$\alpha$ emission is powered by the ionizing radiation from O stars (with typical lifetimes of $\sim 7$ Myr), while the rest-frame UV light receives its major contributions from both O and B stars (i.e., stellar lifetimes up to $\sim 300$ Myr).  The left-hand panel of Figure\ \ref{HaUV_int.fig} visualizes the evolution in $L_{H\alpha}/L_{UV}$ for a set of BC03 stellar populations models with varying star formation histories.  Here, we calculated the intrinsic H$\alpha$ luminosities from the rate of H ionizing photons in the BC03 models.  Applying the recombination coefficients for case B from Hummer \& Storey (1987), for an electron temperature $T_e = 10^4\ K$ and density of $n_e = 10^4\ cm^{-3}$, this gives

\begin {equation}
\log(L_{H\alpha} [erg\ s^{-1}]) = \log( N_{Lyc} [s^{-1}]) - 11.87
\end {equation}

where $N_{Lyc}$ is the production rate of Lyman continuum photons from the stars.  It is noticeable that, except for very young stellar populations or abruptly declining star formation histories, the balance between the intrinsic (i.e., unattenuated) H$\alpha$ and UV emission quickly converges to a value of around -1.7.  We remind the reader that the frequently used Kennicutt (1998) SFR conversions, which assume a 100 Myr old constant star formation population, correspond to the same ratio.  In the following, we adopt $\log(L_{H\alpha}/L_{UV}) = -1.7$ as the expected intrinsic ratio in the absence of dust.  We make the plausible assumption that any observed variation in $L_{H\alpha}/L_{UV}$ will be dominated by extinction effects, and that any (minor) contributions from IMF variations or short-term fluctuations in the star formation history do not depend systematically on $A_V$.  

\begin {figure*}
\plotone{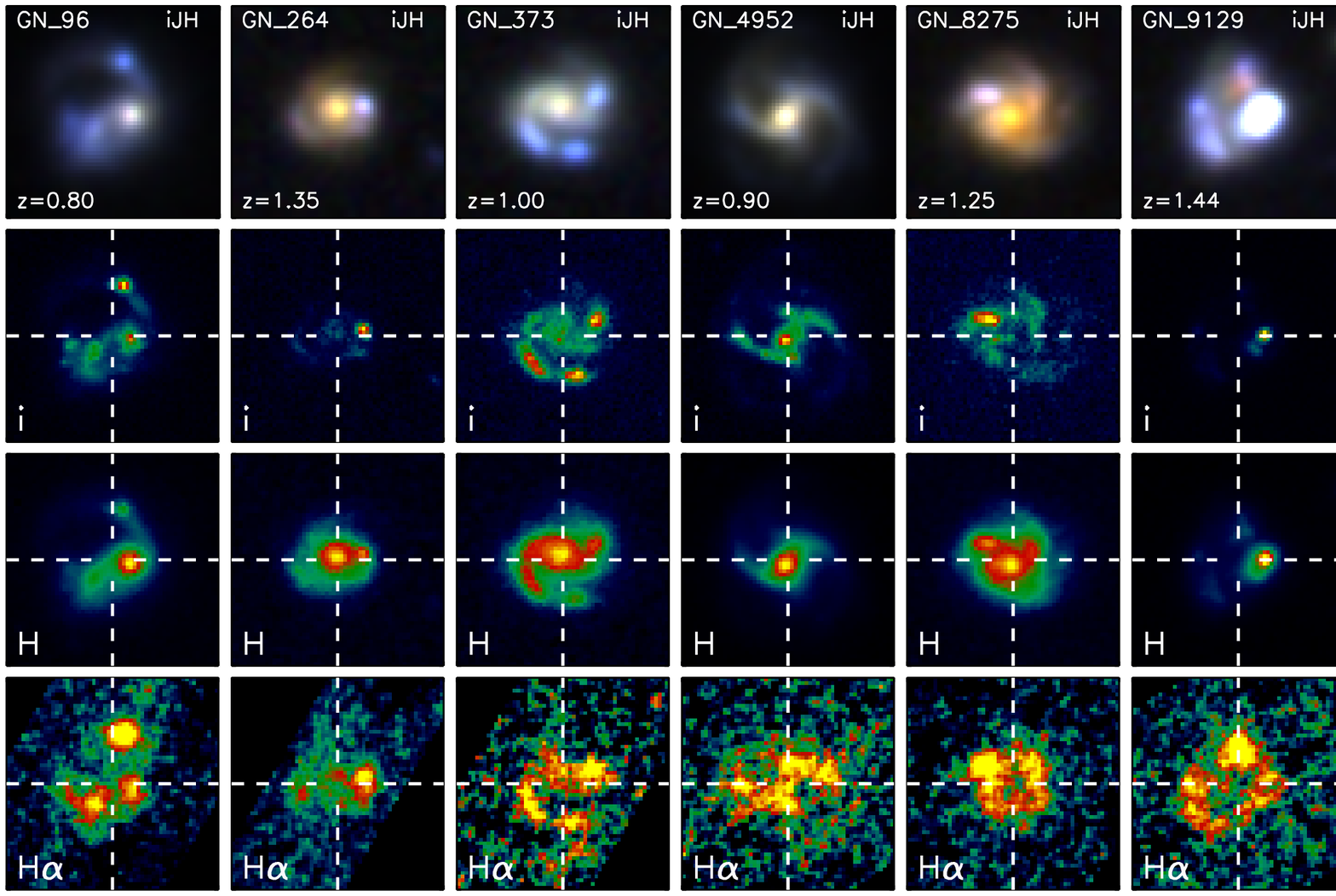}   
\plotone{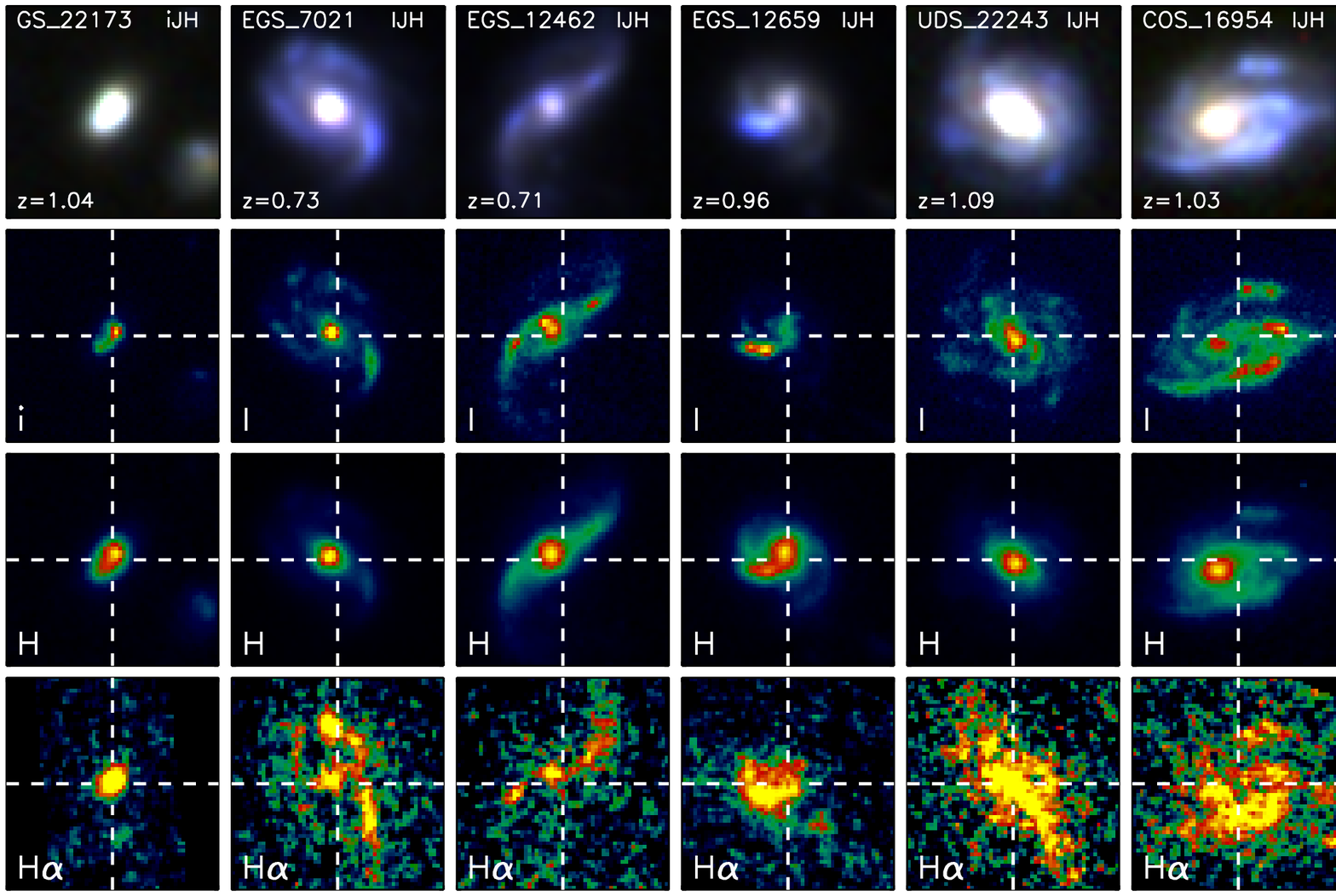}   
\caption{
Gallery of case examples from the massive $z \sim 1$ SFG sample.  PSF-matched 3-color postage stamps sized $3\farcs 4 \times 3\farcs 4$ are composed of $i_{775}J_{125}H_{160}$ for galaxies from the GOODS fields, and $I_{814}J_{125}H_{160}$ for galaxies in EGS/UDS/COSMOS.  Below, we show the surface brightness distributions in $I$, $H$ and H$\alpha$ respectively (at natural resolution, with a slight smoothing applied to the H$\alpha$ maps for visualization purposes).  Blue, star-forming regions present in the $I$ band generally dominate the H$\alpha$ emission as well.  Central peaks in surface brightness (i.e., 'bulges') appear more prominently in the $H$ band.
\label{gallery.fig}}
\end {figure*}

It is then possible to represent different prescriptions for extra extinction towards the nebular regions by lines in a diagram of $L_{H\alpha}/L_{UV}$ versus A$_V$ (right-hand panel of Figure\ \ref{HaUV_int.fig}).  In the absence of extra extinction, the H$\alpha$/UV ratio increases linearly with increasing A$_V$, while it remains relatively constant (in fact drops slightly) with increasing A$_V$ for a so-called 'double Calzetti' prescription.  Neither prescription provides a satisfying description of the locus of observed 3D-HST galaxies in the diagram (red symbols).  Instead, the curve corresponding to Equation\ \ref{bestfit_dustcor.eq} matches the data well.  In Section\ \ref{broad_vs_line.sec}, we will demonstrate that our prescription for H$\alpha$ dust corrections, the associated dependence of H$\alpha$/UV ratios on extinction, and in general the above geometrical picture of a two-component dust distribution still holds on the kiloparsec scale resolution of the HST data.  A direct implication of the observed increase in H$\alpha$/UV ratios with $A_V$ is the fact that H$\alpha$ luminosities are less sensitive to dust than UV luminosities.

Finally, as an additional sanity check of the applicability of our H$\alpha$ dust prescription (Eq.\ \ref{bestfit_dustcor.eq}), we repeated the comparison of SFR estimates and our analysis of the $L_{H\alpha}/L_{UV}$ versus $A_V$ diagram for the upper and lower tertiles of the axial ratio distribution of our galaxies separately.  We confirm that a good correspondence remains, with only marginal systematic shifts.  While the sense of the minor inclination dependence qualitatively agrees with the findings by Wild et al. (2011) for nearby galaxies (edge-on systems requiring relatively less extra extinction because a larger fraction of the obscuring column towards nebular regions is contributed by diffuse dust), we opt not to introduce an extra inclination term in our prescription, given its small amplitude compared to the scatter among galaxies of a given axial ratio.

\section {Case examples}
\label{examples.sec}

Before quantitatively analyzing the H$\alpha$ properties of our sample of massive $z \sim 1$ SFGs, it is insightful to visually inspect a set of case examples.  Figure\ \ref{gallery.fig} shows a subset of 12 SFGs, illustrating the diversity in size, colors and morphologies represented in the sample.  Despite their diverse appearance, some key trends that also characterize the sample as a whole can be noted.

First, the central regions of most (and in particular the larger) galaxies are more pronounced in the observed $H$ than in the $I$ band.  This reflects the negative color gradients (red cores) found in previous work and more compact stellar mass distributions based thereupon (e.g., Szomoru et al. 2011, 2013; Wuyts et al. 2012; Guo et al. 2012).  Second, the $I$-band images appear clumpy or feature spiral signatures more frequently than the rest-optical $H$-band images, echoing findings that the clumps / spiral arms are bluer than the underlying disk (Wuyts et al. 2012; Guo et al. 2012).

New with respect to these previous studies is that we can now also contrast the H$\alpha$ morphologies to the short- and long-wavelength broad-band morphologies.  A visual inspection tells us that overall the H$\alpha$ morphologies resemble more closely the $I$-band images than the $H$-band surface brightness distributions.  Both the rest-UV light and H$\alpha$ emission are dominated by the youngest stellar populations, unlike the rest-optical emission to which the bulk of the stars (including older populations) contribute.  It is however not a trivial observation, because of the different wavelengths probed and dust effects discussed in Section\ \ref{Halpha_dust.sec}.  Moreover, exceptions exist in which the H$\alpha$ morphology does not trace the $I$-band light.  For example, the $I$-band emission of GN\_9129 is entirely dominated by a clump that is not the most prominent H$\alpha$ clump in the object.  In addition, for the smaller galaxies we are hard-pressed to make a statement on whether the visual morphological match of the H$\alpha$ map is closer to $I$ than to the $H$ band.  Galaxies such as GS\_22173 just appear compact in all bands considered.  These results are reminiscent of the conclusions drawn by Nelson et al. (2012) on the basis of a sample of 56 equivalent width selected galaxies in 3D-HST.

\section {Results}
\label{results.sec}

\subsection {H$\alpha$ emission traces better the rest-UV than the rest-optical light}
\label{crosscor.sec}

\begin{figure}[t]
\plotone{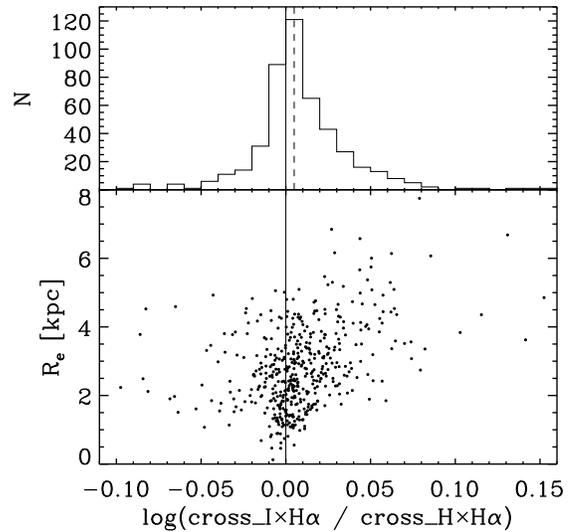}   
\caption{
{\it Top:} Histogram of the ratio of cross-correlation strengths between the H$\alpha$ and $I$-band surface brightness distribution and the H$\alpha$ and $H$-band surface brightness distribution respectively.  The vertical dashed line marks the median.  The distribution is asymmetric, with a tail towards positive values of $\log(cross\_I / cross\_H)$, representing objects whose H$\alpha$ profiles resemble more those observed in the $I$ band than in the $H$ band.  {\it Bottom:} The closer correspondence between the H$\alpha$ and $I$-band images is most prominent for the larger galaxies in our sample (with $R_e$ being the effective radius measured by fitting a Sersic profile to the $H$-band light profile).  More compact galaxies generally look similar at all considered wavelengths ($I$, $H$, and H$\alpha$).
\label{crosscor.fig}}
\end{figure}

\begin {figure*}
\plottwo{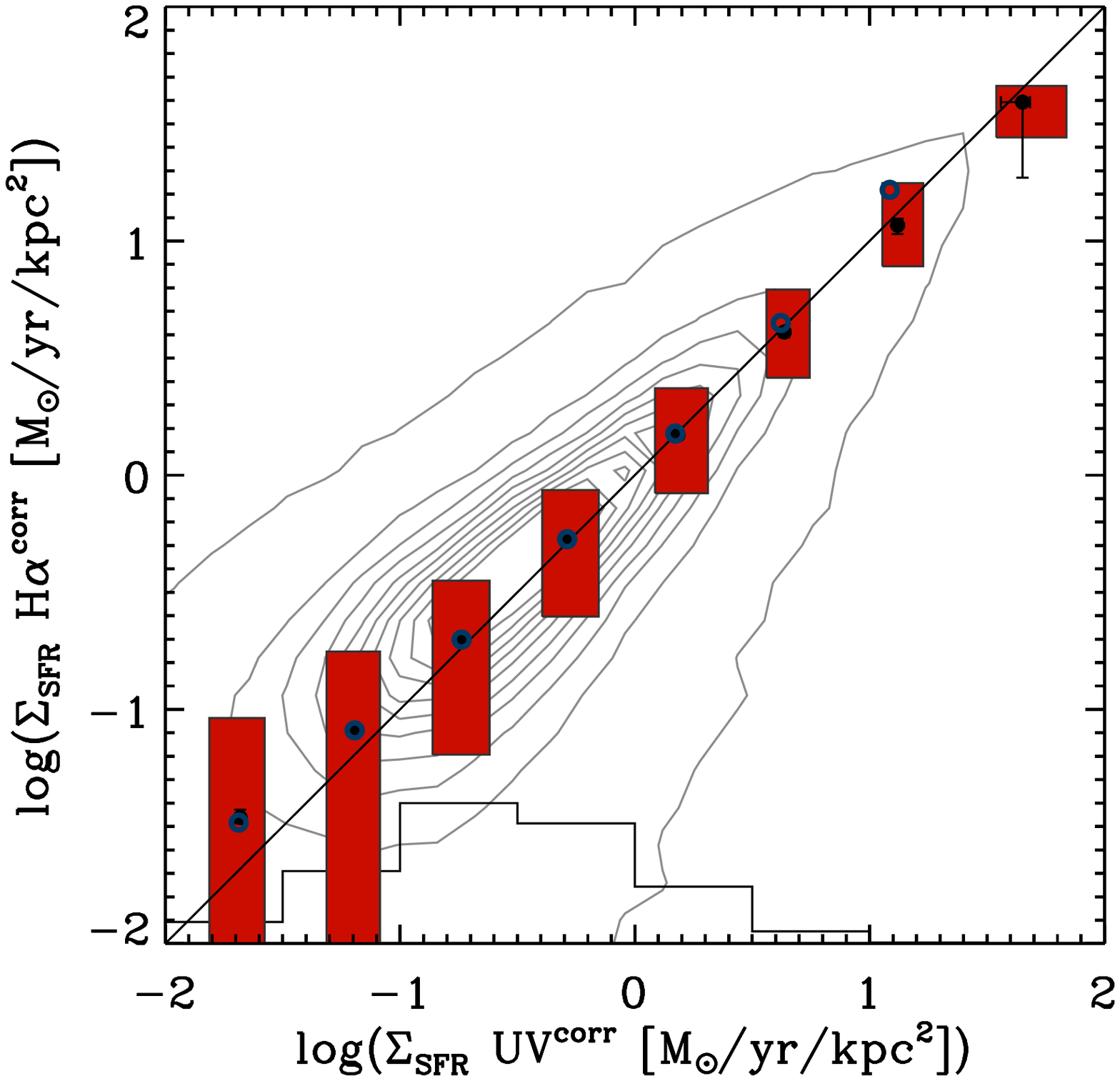}{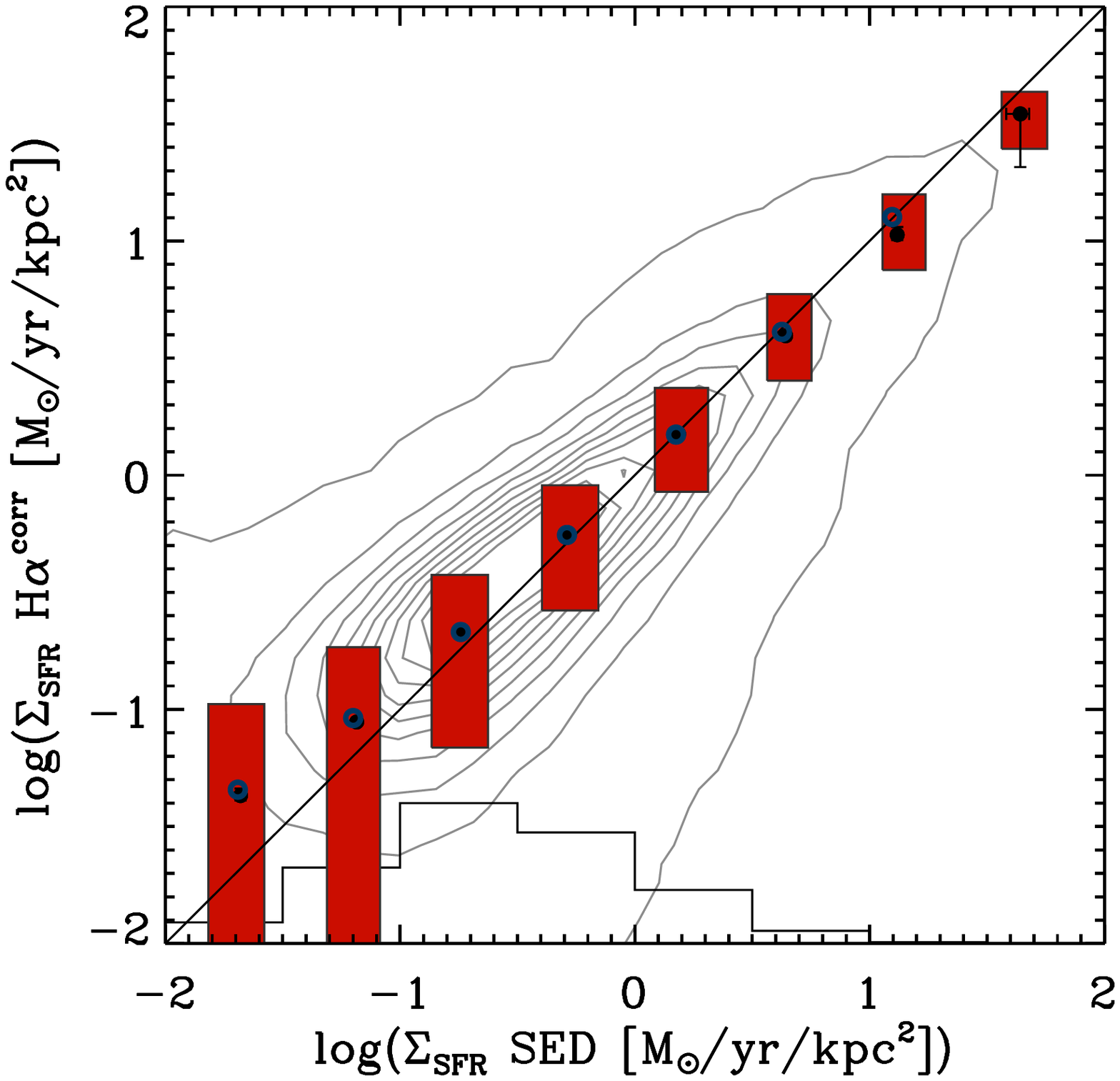}   
\plottwo{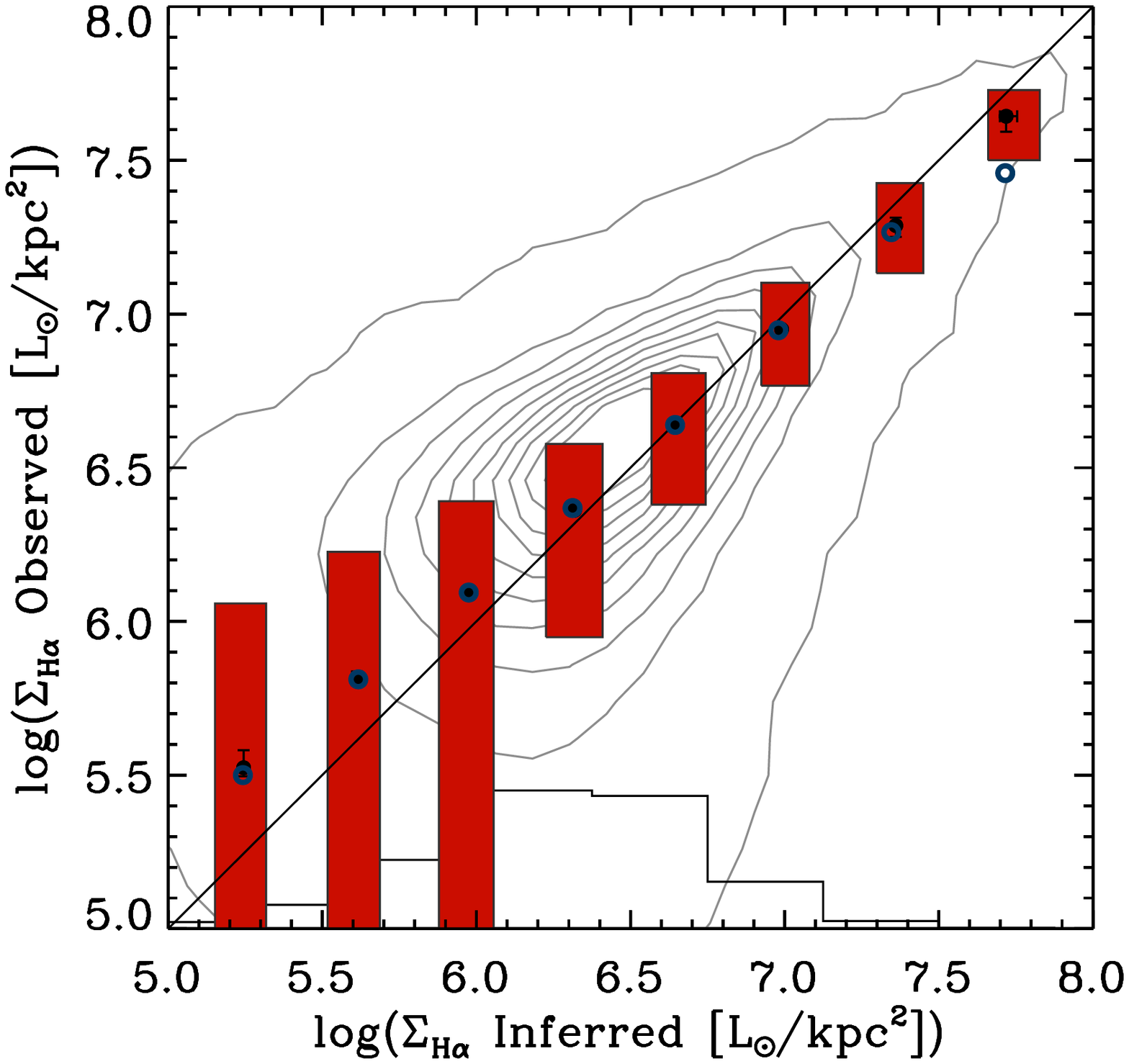}{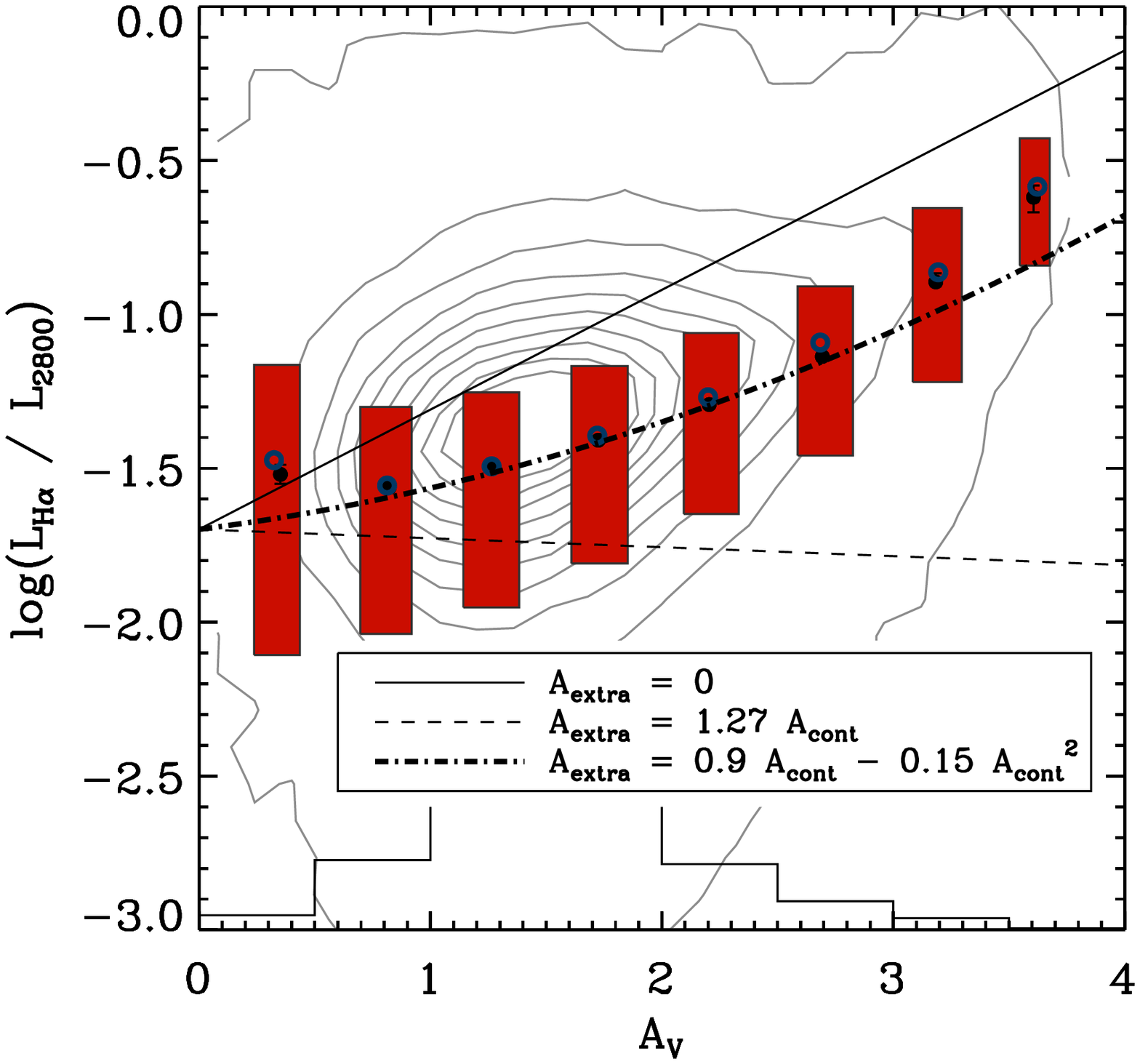}   
\caption{
Comparison of resolved galaxy properties based on H$\alpha$ versus resolved stellar population properties derived from broad-band information alone.  Contours indicate the locus of all spatial bins of the galaxies in our sample.  Red bars with black dots mark the central 50th percentile and median, respectively.  With open blue circles, we mark the median results for the subset of galaxies in the GOODS fields, where the high-resolution multi-wavelength sampling is best, with 7 bands.  Black histograms illustrate the distribution of spatial bins.  The dust corrected H$\alpha$ SFRs of individual spatial bins within galaxies compare favorably to estimates from the UV or SED modeling ({\it top panels}).  {\it Bottom left:} A good correlation remains when plotting the directly observed H$\alpha$ luminosity surface density (i.e., without dust correction) as a function of what would be inferred on the basis of broad-band information, demonstrating the $\Sigma_{SFR}$ agreement is not just due to the appearance of $A_V$ in the quantities on both axes.  {\it Bottom right:} The H$\alpha$/UV luminosity ratio of individual spatial bins within galaxies shows a similar dependence on visual extinction as observed in a galaxy-integrated sense in Figure\ \ref{HaUV_int.fig}.
\label{broad_vs_line.fig}}
\end {figure*}

In order to quantify the above visual impressions, we analyze the morphological match between H$\alpha$ and $H$ and H$\alpha$ and $I$ band surface brightness distributions respectively by cross-correlating the images.  In doing so, we allow minor shifts of the galaxy centroid before mapping the grism image to the astrometric frame of the CANDELS broad-band images.  This accounts for uncertainties in the position of the grism spectra on the detector and degeneracies between on the one hand the systemic redshift and on the other hand morphological k-corrections between the H$\alpha$ line map and the galaxy's F140W light distribution that was used as template in the redshift determination.  For each band, we adopted the shift leading to the largest correlation coefficient, in most cases limited to 0 - 3 pixels\footnote{One pixel corresponds to 23\AA\  in wavelength, and $0\farcs06$ spatially.}.  The ratio of $I$-to-$H$ band correlation coefficients then serves as a quantitative measure of how much better the H$\alpha$ morphology corresponds to the $I$-band light distribution compared to the $H$ band.  Figure\ \ref{crosscor.fig} demonstrates what we inferred by eye: a majority (65\%) of galaxies in our sample shows a stronger cross-correlation between the $I$ and H$\alpha$ morphologies than between the $H$ and H$\alpha$ morphologies.  This difference is most pronounced among the more extended systems.  For galaxies with semi-major axis lengths larger than 3 kpc (as measured with GALFIT by van der Wel et al. 2012), more than 75\% shows a better match between H$\alpha$ and the $I$ band than between H$\alpha$ and the $H$ band.  GN\_96, GN\_373 and GN\_4952 in Figure\ \ref{gallery.fig} serve as examples of 3 - 5 kpc sized sources with $\log(\rm cross\_I\times H\alpha\ /\ \rm cross\_H\times H\alpha) \sim 0.05 - 0.07$.  For the largest star-forming galaxies at $z \sim 1$, with $R_e > 5$ kpc, this fraction increases further to above 90\% (see, e.g., EGS\_7021 and COS\_16954 in Figure\ \ref{gallery.fig}).  We stress that the above statements on the resemblance between line emission maps and broad-band images of different wavelengths are of a comparative nature.  Substantial pixel-to-pixel variations exist in the flux ratio between H$\alpha$ and any single broad band (even at short wavelengths), preventing a simple scaling from a monochromatic broad-band image to H$\alpha$ without additional information.  In the next Section, we explore how multi-wavelength broad-band information improves this situation, and furthermore allows us to dust correct H$\alpha$ maps to star formation maps.

\subsection{Comparing resolved broad-band and emission line diagnostics}
\label{broad_vs_line.sec}

\subsubsection{Resolved dust and SFR calibrations}

We now proceed to translate the direct observables (i.e., surface brightness profiles in H$\alpha$ and multiple broad bands) to more physically relevant quantities, following the resolved SED modeling and H$\alpha$ extinction correction methodologies outlined in Section\ \ref{methodology.sec}.  Figure\ \ref{broad_vs_line.fig} contrasts various H$\alpha$-based diagnostics to stellar population properties inferred using broad-band information alone.  We place each spatial bin of each galaxy in the respective diagrams, and consider the locus of observed points, marked by contours.  Due to noise, some individual spatial bins have negative H$\alpha$ fluxes and therefore do not appear in the plot.  These pixel bins are however included in the binned median and central 50th percentile statistics marked in red.  In a resolved analogy to Figure\ \ref{HaUV_int.fig}, we confirm the distribution of observed H$\alpha$/UV ratios to relate to the visual extinction $A_V$ inferred from resolved SED modeling as would be expected from Equation\ \ref{bestfit_dustcor.eq}.  In contrast, $A_{extra}=0$ or $A_{extra}=1.27A_{cont}$ do not reproduce the observed locus of galaxy spatial bins in the diagram.  Consequently, it comes as no surprise that, when applying dust corrections according to Equation\ \ref{bestfit_dustcor.eq} to the H$\alpha$ emission, we find the H$\alpha$-based $\Sigma_{SFR}$ to correspond well to dust-corrected UV-based estimates.  To compute the latter, we converted the rest-frame 2800\AA\ luminosity  $L_{2800}$  to an unobscured UV SFR based upon the Kennicutt (1998) conversion $SFR_{UV,\ uncorrected} = 3.6 \times 10^{-10} L_{2800}/L_{\sun}$, and finally applied an extinction correction based on the broad-band SED modeling which assumed a Calzetti et al. (2000) attenuation law.  Since the visual extinction appears in the quantities on both axes of the H$\alpha$- versus UV-based dust-corrected $\Sigma_{SFR}$, we also show that a strong correlation remains when plotting the directly observed H$\alpha$ luminosity surface density (i.e., uncorrected for dust) as a function of it would be inferred to be on the basis of the available broad-band information.

\subsubsection{The resolved main sequence of star formation}
\label{resMS.sec}
\begin{figure}
\plotone{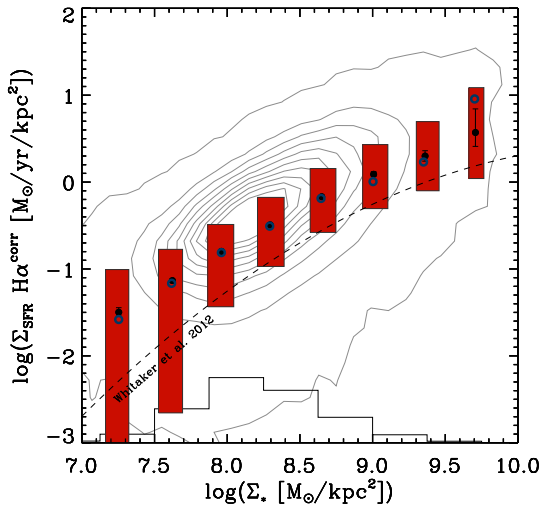}   
\caption{
Equivalent to the so-called 'main sequence of star formation' for galaxies, resolved subregions within galaxies exhibit a correlation between the level of ongoing star formation and the surface density of assembled stellar mass (plot style as in Figure\ \ref{broad_vs_line.fig}).  For reference, the dashed line marks the slope of the galaxy-integrated main sequence from Whitaker et al. (2012).
\label{resMS.fig}}
\end{figure}

Exploiting our confidence gained by the analysis of SFR estimates and dust treatment, we now turn to Figure\ \ref{resMS.fig} that contrasts the H$\alpha$-based SFR surface density of individual sub-galactic regions to the surface density of stellar mass present at the respective location in the galaxy.  As such, the diagram can be regarded as a spatially resolved version of the SFR-mass plane that has been the focus of many galaxy evolution studies in recent years.  We find a relation between the star formation activity and amount of assembled stellar mass to be present for sub-regions within galaxies, as is the case for galaxies as integrated units.  Similar to the well-established, galaxy-integrated main sequence of star formation, we observe a near-linear slope:

\begin {equation}
\log(\Sigma_{SFR} [M_{\sun}\ yr^{-1}\ kpc^{-2}]) = -8.4 + 0.95 \log(\Sigma_{*} [M_{\sun}\ kpc^{-2}]),
\end {equation}

as fit to the spatial bins with $\log (\Sigma_*) < 8.8$.
Akin to galaxy-integrated observations at the high-mass end by, e.g., Whitaker et al. (2012) , the 'resolved main sequence' shows a tendency to flatten at the high stellar surface mass density end.  Consequently, a linear fit to all spatial bins, spanning the entire range in $\Sigma_*$, yields a slightly shallower slope, of 0.84.  We conclude that, even if global galaxy properties such as halo mass or cosmological accretion rate drive the stellar build-up within galaxies, this happens in such a way that on average the ongoing and past-integrated star formation activity track each other on sub-galactic as well as galactic scales.  This is at least the case down to the $\sim 1$ kpc scales probed at the WFC3 resolution, albeit with a significant scatter, and possible exceptions (i.e., reduced specific star formation rates) at the highest stellar surface mass density end.

In the remainder of the paper, we will investigate the deviation from a perfect $\Sigma_{SFR} \propto \Sigma_*$ relation more closely, and illustrate where spatially within galaxies variations in local specific SFR occur.  In analogy with the galaxy-integrated main sequence of star formation, the clumps/spiral arms featuring excess star formation with respect to the smooth underlying disk (Section\ \ref{clump_nature.sec}) can be thought of as subgalactic equivalents to the starbursting outliers above the main sequence.

\subsection{Resolved H$\alpha$ properties in the two-dimensional profile space}
\label{prof2D.sec}

\subsubsection{The nature of excess surface brightness regions}
\label{clump_nature.sec}

\begin{figure*}[t]
\plottwo{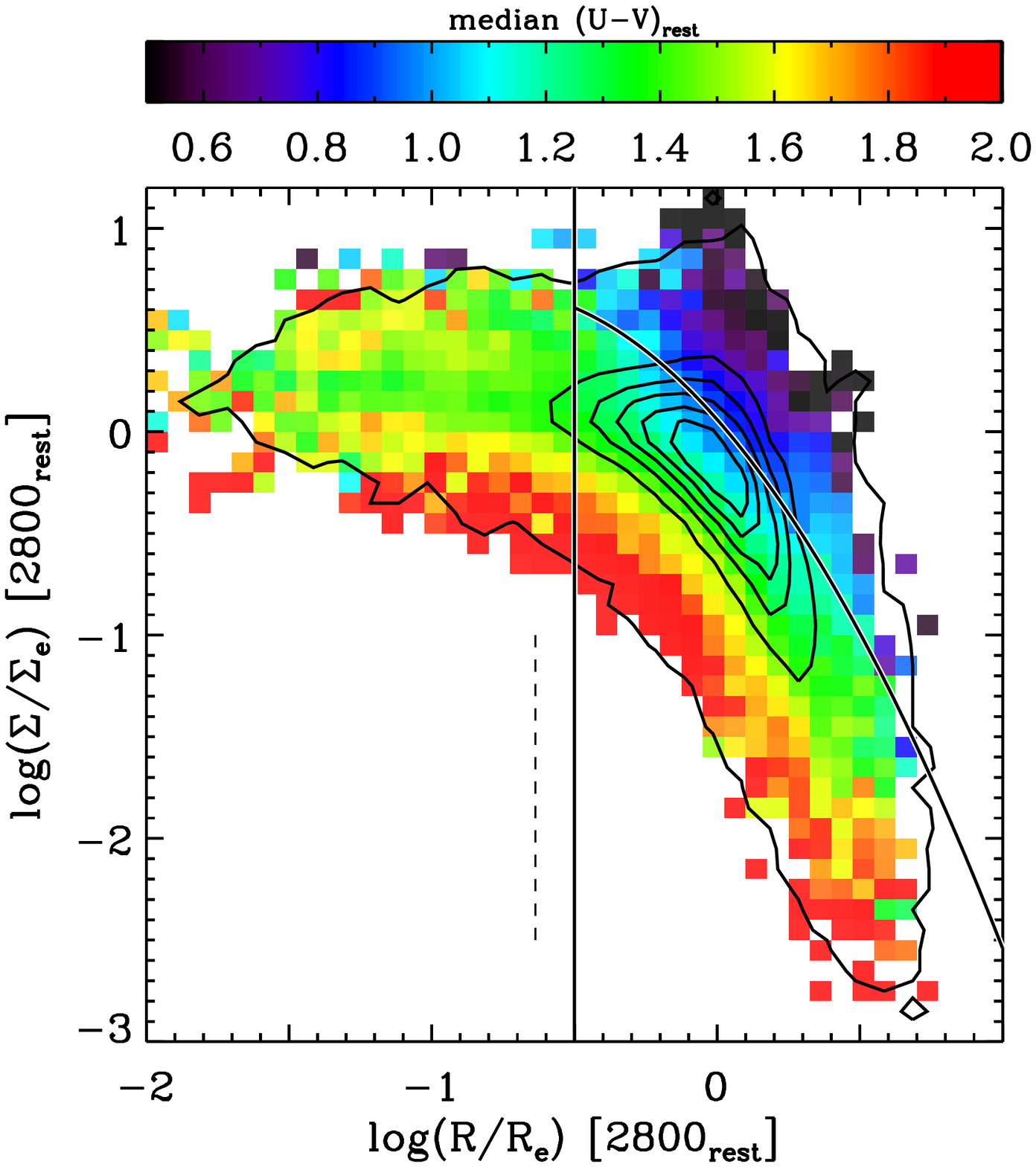}{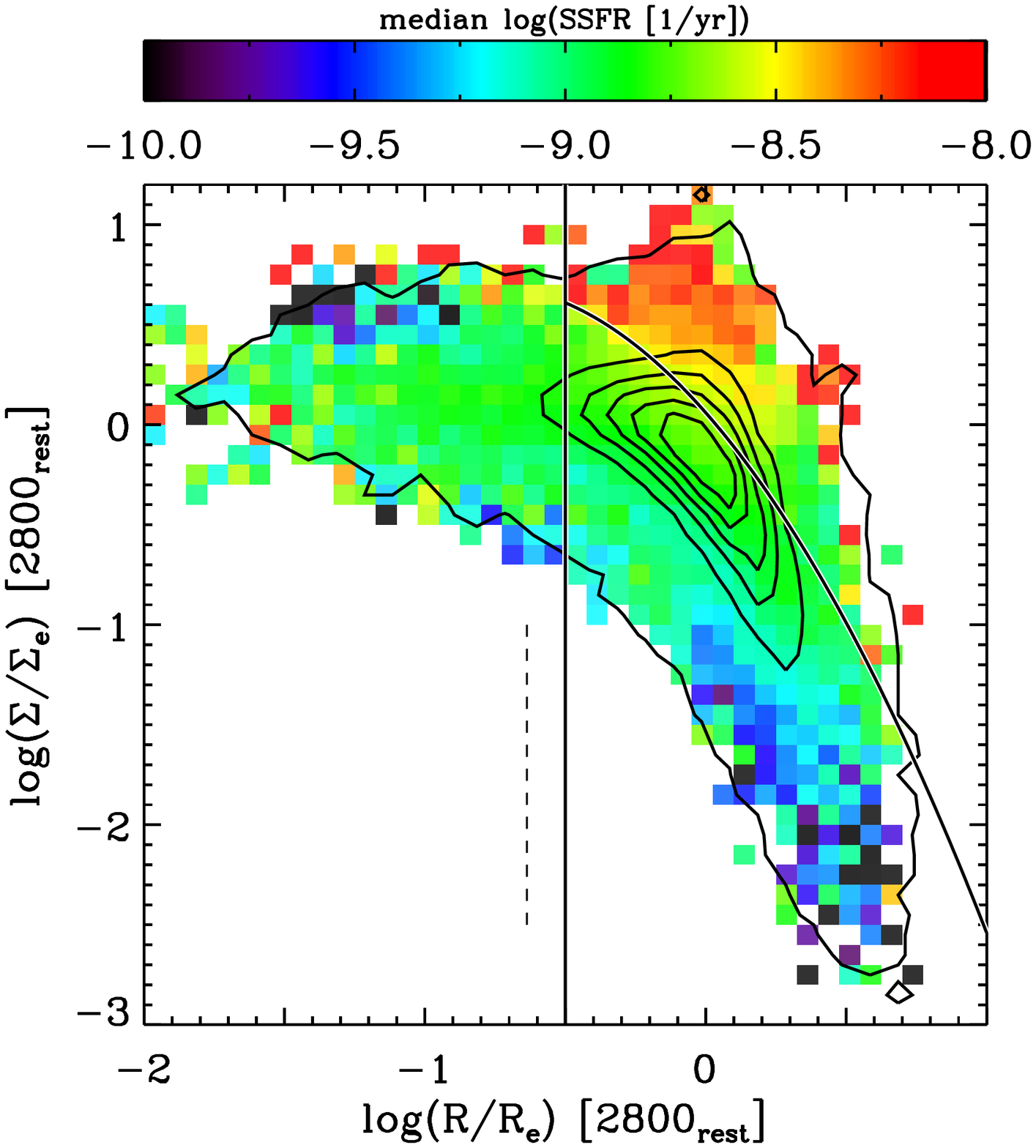}   
\plottwo{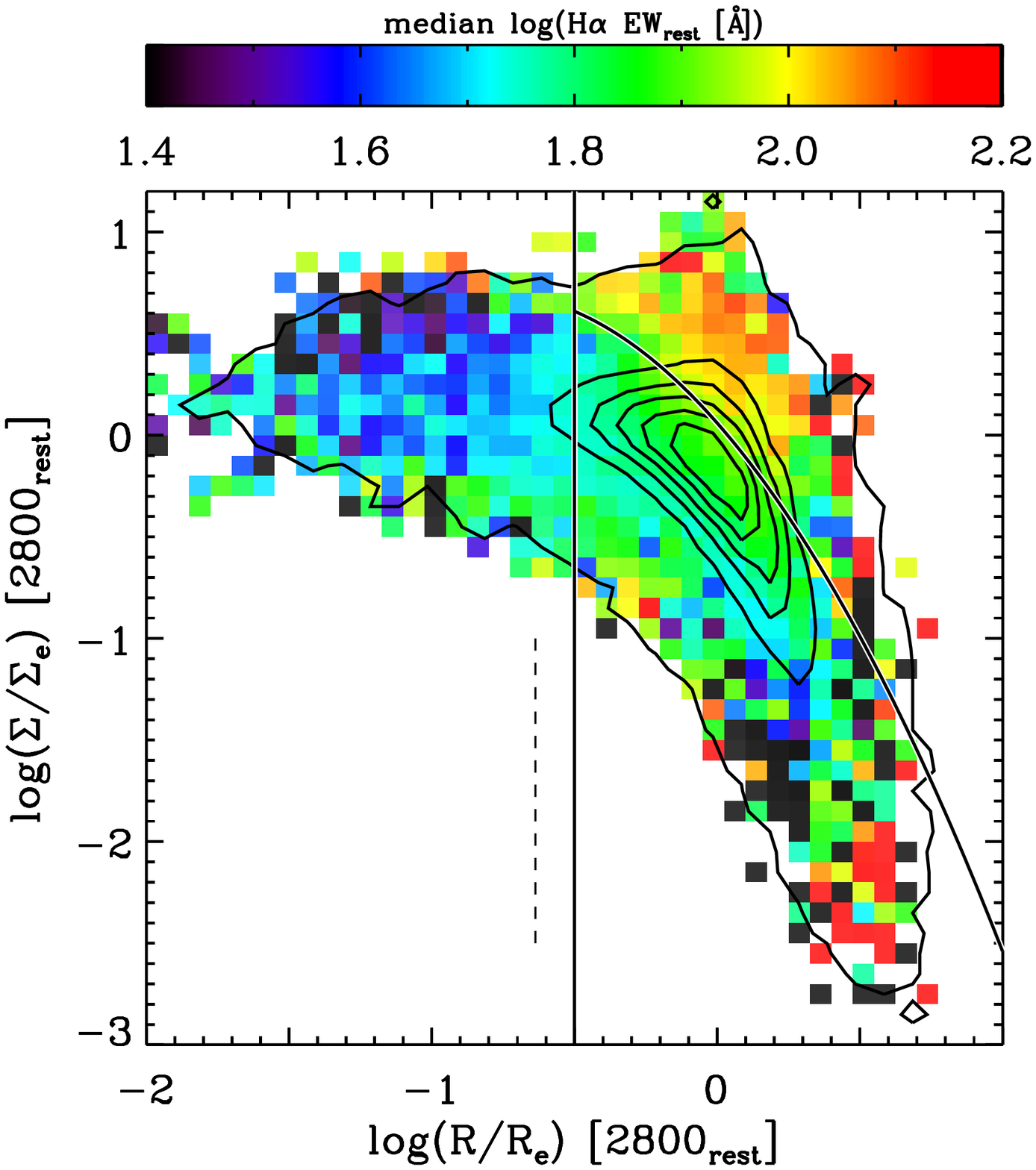}{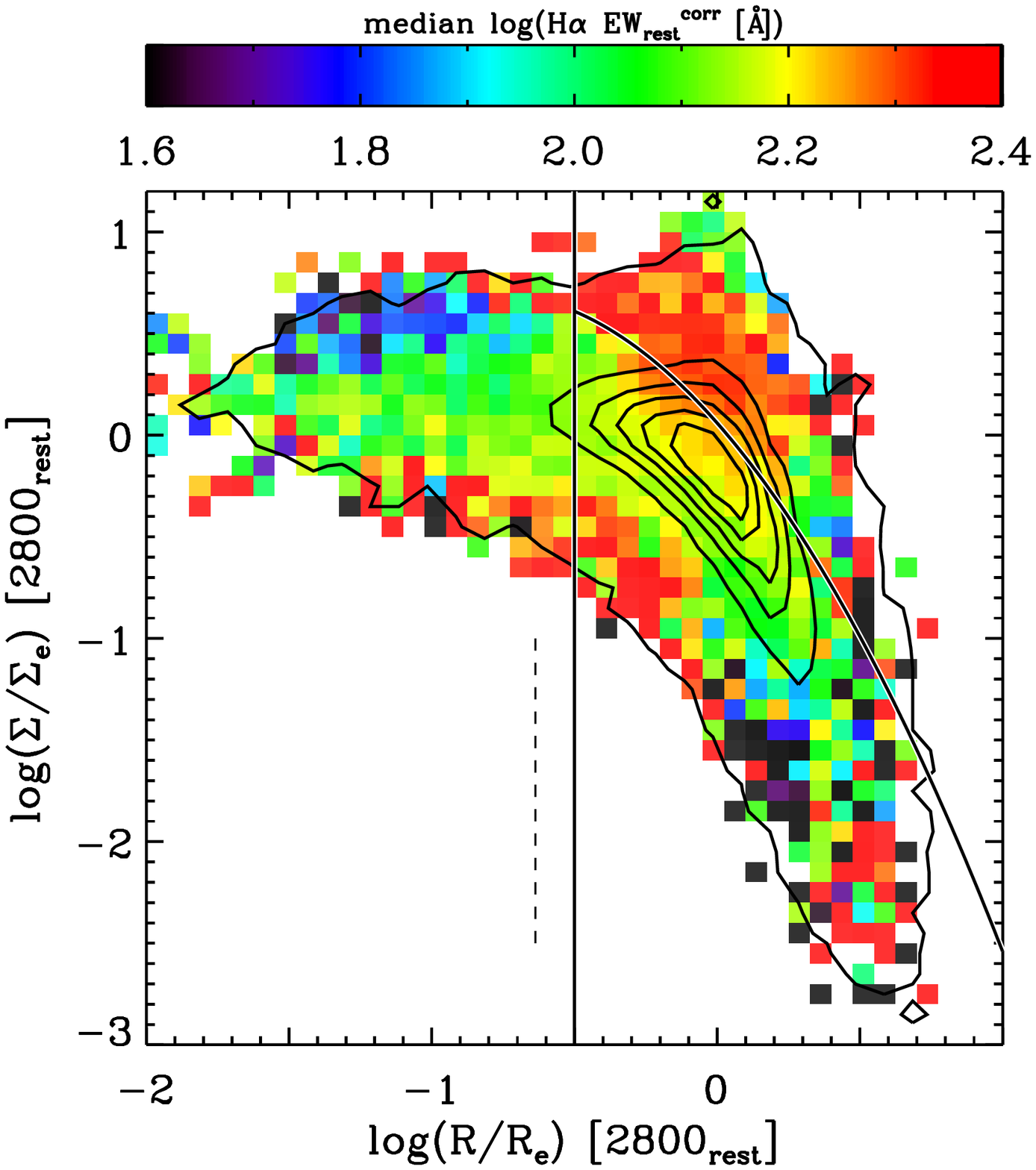}   
\caption{
Co-added normalized rest-UV surface brightness profile of the SFGs in our sample.  The vertical dashed line indicates the typical resolution.  Contours denote the density of spatial bins.  The top panels are color coded by purely broad-band based quantities, namely the rest-frame optical color $(U-V)_{rest}$ ({\it top left}) and the specific star formation rate inferred from resolved stellar population modeling ({\it top right}).  The color coding in the bottom panels marks the H$\alpha$ rest-frame equivalent width as observed directly ({\it bottom left}) and after correction for extra extinction towards the nebular regions ({\it bottom right}).  Solid lines separate the central, outer disk, and clump/spiral arm regime.  Spatial bins in the clump/spiral arm regime are characterized by bluer broad-band colors and higher $H\alpha$ EWs, consistent with the elevated SSFR with respect to the underlying disk inferred from the resolved SED modeling.
\label{clump_nature.fig}}
\end{figure*}

With the resolved H$\alpha$ properties in hand, we now revisit the conclusion by Wuyts et al. (2012) based on broad-band SED modeling that clumps in massive high-redshift SFGs correspond to star-forming phenomena: regions that exhibit an enhanced level of star formation with respect to the underlying disk.  Given short timescales for these spatial fluctuations in local star formation, the same substructure is not necessarily present in the distribution of stellar mass, as would be expected if the observed irregular morphologies were all due to merging activity.

Following Wuyts et al. (2012), we employ co-added normalized profiles to infer the characteristic star formation properties of pixels belonging to the galaxy's central regions, off-center clumps (or spiral arms), and the (interclump) outer disk (Figure\ \ref{clump_nature.fig}).  In each of these diagrams, we consider all spatial bins of all 473 SFGs as being part of one meta-galaxy, representing the massive SFG sample as an ensemble, whose resolved stellar populations we then dissect.  We apply identical criteria to Wuyts et al. (2012) in order to separate spatial bins belonging to the three regimes:

\begin{align}
[center]\ \ &  \log(R/R_e) < -0.5 & \label{center.eq} \\
[disk]\ \ &  \log(R/R_e) > -0.5\ \& & \label{disk.eq} \\
 & \log(\Sigma/\Sigma_e) < 0.06 - 1.6 \log(R/R_e) - \log(R/R_e)^2 & \nonumber \\
[clump]\ \ & \log(R/R_e) > -0.5\ \& & \label{clump.eq} \\
 & \log(\Sigma/\Sigma_e) > 0.06 - 1.6 \log(R/R_e) - \log(R/R_e)^2. & \nonumber
\end{align}

where $\Sigma/\Sigma_e$ is the rest-frame 2800\AA\ surface brightness normalized by the average surface brightness level within the effective radius $R_e$.  As in Wuyts et al. (2012), we define radii on elliptical apertures whose center and axial ratio are derived from the stellar mass maps.  Since the distance from the center of a spatial bin to the center of mass of the galaxy can in principle be arbitrarily small, we note that the precise galactocentric radii at small $\log(R/R_e) \ll -0.5$ becomes meaningless.  Throughout the paper, we therefore treat all pixels with $\log(R/R_e) < -0.5$ as belonging to the inner regions of galaxies, without further regard to their precise value of $R/R_e$.  In the top panels of Figure\ \ref{clump_nature.fig}, we first demonstrate that the conclusions drawn by Wuyts et al. (2012) based on broad-band information alone also apply to our H$\alpha$ sample at $0.7<z<1.5$.  Off-center clumps identified at short wavelengths by an excess in surface brightness are characterized by bluer rest-optical colors than the underlying disk.  Taking one step further away from the direct observables, stellar population synthesis modeling of the multi-wavelength HST colors implies they correspond to sites of enhanced star formation activity (SSFR) with respect to the underlying disk.

The H$\alpha$ equivalent width can be considered as a complementary, less model-dependent diagnostic of the star formation activity.  It is defined as the ratio of the star formation sensitive line emission strength over the continuum flux, which probes the bulk of the stars present in the galaxy.  As the emission line and continuum are measured at the same wavelength, and both components are subject to the attenuating column of diffuse dust in the interstellar medium of the galaxy, extinction effects due to the latter divide out.  We observe the H$\alpha$ equivalent width to be higher in the clumps/spiral arm regions (with a median rest-frame EW$_{H\alpha} \approx 100\AA$) than in the underlying smooth disk (EW$_{H\alpha} \sim 70\AA$), while the lowest equivalent widths are measured at small radii.  The latter are associated with the regions of highest stellar surface mass density, and are related to the flattening slope of the resolved main sequence at high $\Sigma_*$ (see Section\ \ref{resMS.sec}).  The implications of the radial trends will be discussed in depth in a forthcoming comprehensive overview of H$\alpha$ and star formation profiles as a function of galaxy mass and SFR by E. Nelson et al. (in prep).  Here, we consider in more detail the characteristic trends with surface brightness at a fixed radius.  As discussed in Section\ \ref{calibrating.sec}, an additional extinction correction to account for the obscuring material in birth clouds from which the H$\alpha$ line emission originates is required for the equivalent width to be an unbiased proxy of the star formation activity.  In the bottom right panel of Figure\ \ref{clump_nature.fig}, we apply this correction and find an excess in the clumps' intrinsic H$\alpha$ equivalent width by 25\% in the median with respect to the underlying disk (i.e., interclump regions) at the same galactocentric radii.  We note that the median excess between the equivalent width of clump and disk pixels does not reflect the full dynamic range in rest-2800\AA\ surface brightness variations at a given radius and the extent of the associated trend in the H$\alpha$ equivalent width.  The reason is that the distribution of disk and especially clump pixels is dominated in numbers by pixels that lie close to the dividing line between the two regimes (see contours).  The color coding in Figure\ \ref{clump_nature.fig} illustrates that excess equivalent widths by factors of 2 to 3 are commonly observed when considering the full range of surface brightness levels observed at a given radius.

\begin{figure}
\plotone{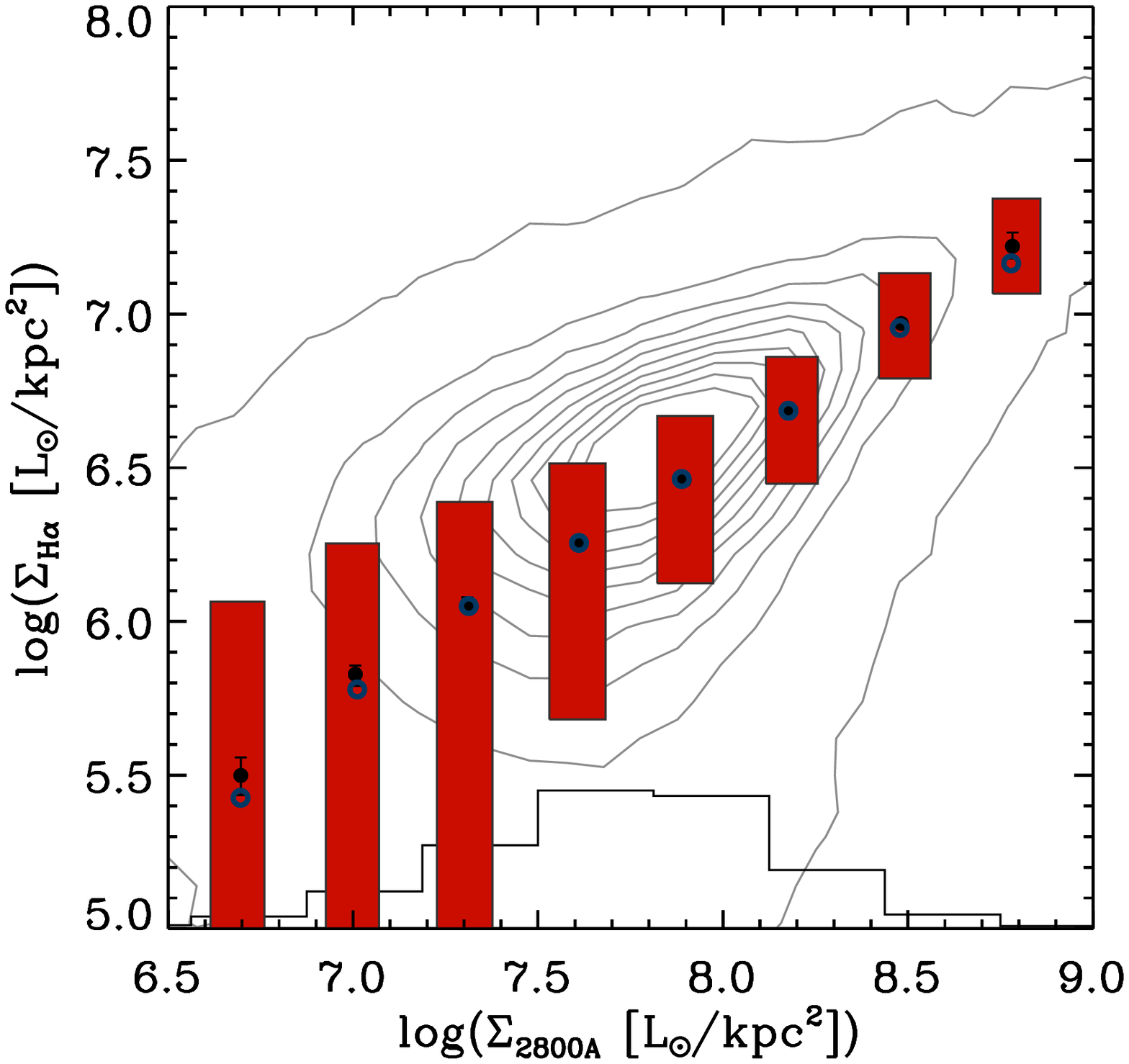}   
\plotone{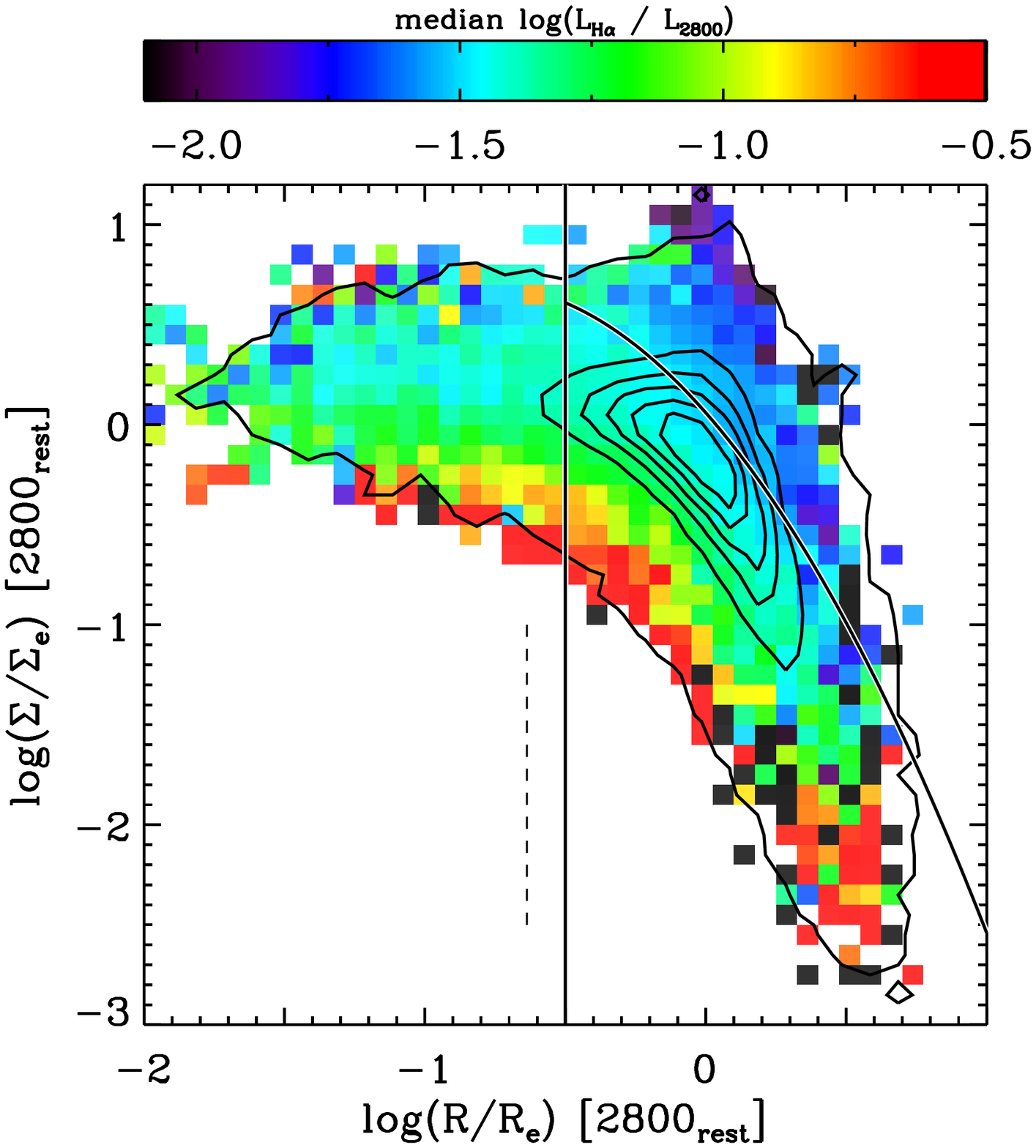}   
\caption{While the H$\alpha$ luminosity surface density correlates with the rest-frame 2800\AA\ luminosity surface density of individual bins ({\it top panel}), variations in the H$\alpha$/UV luminosity ratio are present and show a systematic dependence on the location of pixels spatially within the galaxies ({\it bottom panel}, identical in style to Figure\ \ref{clump_nature.fig}).  The reduced H$\alpha$/UV luminosity ratios of clumps with respect to the fainter interclump regions may arise from spatial variations in the effective attenuation (i.e., reduced dust extinction towards those sites of enhanced star formation where the rest-UV surface brightness is observed to peak).  
\label{HaUV_prof.fig}}
\end{figure}

In order to address whether the clump and spiral arm regions selected through our methodology differ in other aspects than star formation activity from the diffuse outer disk, we now investigate the relation between the H$\alpha$ and rest-2800\AA\ luminosity surface densities (both uncorrected for extinction; Figure\ \ref{HaUV_prof.fig}, top panel).  As already hinted at by the similar morphological appearance of our galaxies in H$\alpha$ and ACS imaging (Section\ \ref{examples.sec} \&\ \ref{crosscor.sec}), the UV and H$\alpha$ SFR tracers correlate strongly, both spatially and in amplitude.  Their measured luminosity surface densities prior to any dust correction, however, do not follow a relation with a unity slope.  Instead, the formal best-fit linear relation is
\begin {equation}
\log(\Sigma_{H\alpha}\ [L_{\sun}/kpc^2]) = 0.2 + 0.8 \log(\Sigma_{2800}\ [L_{\sun}/kpc^2]).
\end {equation}
Two potential contributors to the deviation from a unity slope, scatter around the relation, and in more general terms behind variations in H$\alpha$/UV luminosity ratios, are dust extinction and variations in the star formation history (see Section\ \ref{Halpha_dust.sec}).  While both H$\alpha$ and UV light are tracing young stellar populations, they probe recent star formation over different timescales (around 10 Myr and 100 Myr, respectively).  The UV and H$\alpha$ radiation are furthermore emitted at different wavelengths and moreover continuum and nebular emission may travel different paths through the galaxies' dusty ISM.  The combination of these effects leads to systematically larger H$\alpha$/UV ratios with increasing extinction compared to the intrinsic (i.e., dust-free) ratio.  

\begin{figure*}[t]
\plottwo{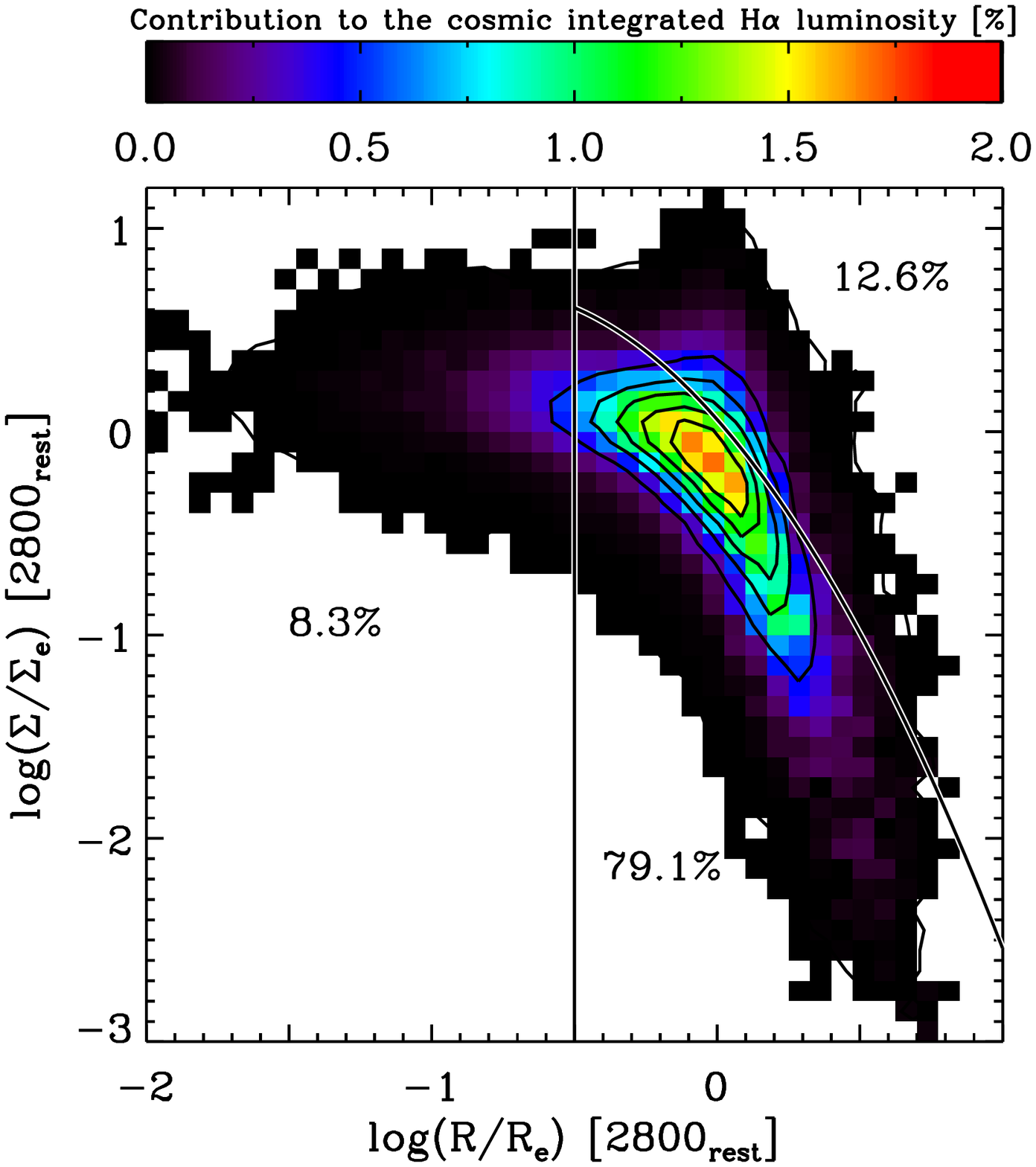}{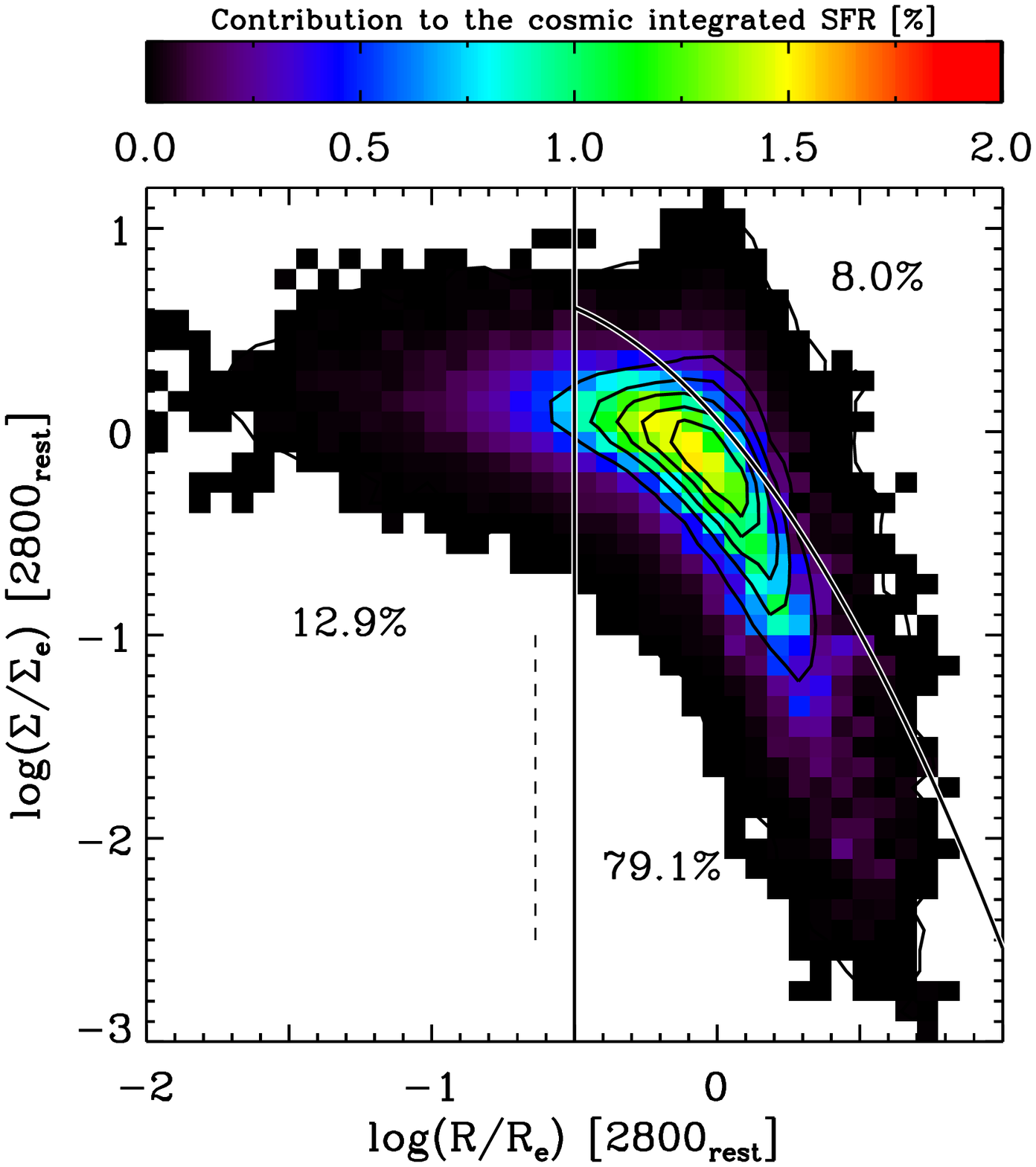}   
\caption{
Co-added normalized rest-UV surface brightness profile of the SFGs in our sample (identical in style to Figure\ \ref{clump_nature.fig}).  The color coding of each bin in the diagram indicates the percentage of the total integrated H$\alpha$ luminosity ({\it left panel}) or dust-corrected SFR based thereupon ({\it right panel}) summed over all massive SFGs in our sample, originating from the respective region in 2D profile space (i.e., galactocentric radius and rest-2800\AA\ surface brightness level).  Differences between the fractional contribution to the cosmically integrated H$\alpha$ luminosity and SFR arise from inhomogeneous dust distributions within galaxies: thicker dust columns towards the center and slightly reduced obscuration to the rest-UV selected clumps.  The vast majority of new-born stars in $z \sim 1$ galaxies are being formed in the disk.
\label{clump_budget.fig}}
\end{figure*}

Evaluating the spatial variation of H$\alpha$/UV ratios within our $z \sim 1$ SFGs (Figure\ \ref{HaUV_prof.fig}, bottom panel), we find the clumps/spiral features to correspond to lower $L_{H\alpha}/L_{UV}$ than the underlying disk at the same radius.  Three effects can potentially contribute to this trend.  The first relates to measurement uncertainties, rather than physics.  If the true H$\alpha$/UV ratio were constant across galaxies, measurement errors would preferentially push those spatial bins over the clump selection threshold for which $\Sigma_{2800}$ is overestimated.  Since the 2800\AA\ luminosity appears in the denominator, the measured H$\alpha$/UV ratio for those bins would consequently be lower than the true value.  While possibly contributing in a minor way, this effect is unlikely to be by itself responsible for the observed trend, given the small uncertainty on $\Sigma_{2800}$ (see Appendix) and the fact that the clumps are also confirmed to be physically distinct from the underlying disk in quantities that are measured independently of the 2800\AA\ luminosity (e.g., the H$\alpha$ equivalent width).

A more physical interpretation is that the excess surface brightness regions selected from rest-UV ACS imaging are not only regions characterized by enhanced star formation, but preferentially those local enhancements in star formation that are the least obscured.  Speculatively, outflows driven by the enhanced star formation density in 'clumps' (see, e.g., Newman et al. 2012) may be responsible for the clearing of gas and dust towards these sightlines.\footnote{See also Calzetti et al. (2005) for a discussion on the increased efficiency of dust and gas clearing from high star formation density regions, such as starbursts, compared to normal disk galaxies.}

Finally, we cannot exclude that recent variations in the star formation history contribute to the lower H$\alpha$/UV ratios of clumps as well.  We note, however, that a recent upturn of the SFR in spatial bins that are classified as clumps or spiral features would lead to elevated H$\alpha$/UV ratios (i.e., relatively more O stars compared to the number of B stars), the opposite of what is observed.  Interpreting the H$\alpha$/UV ratios in terms of changes in the star formation history would therefore invoke a rapidly declining star formation rate (see Figure\ \ref{HaUV_int.fig}), and consequently a prior (possibly very short-lived) phase of even more intense star formation.  The latter phase necessarily would have to be embedded, since otherwise it would itself be picked up by the UV clump selection.  As such, both physical explanations could go hand in hand, in a scenario where outflows quickly clear the gas and dust from clumps, simultaneously reducing their star formation activity and obscuration, and consequently lowering the observed H$\alpha$/UV ratio.  If true, this would imply that resolved observations at longer wavelengths with ALMA or PdBI/NOEMA may reveal new/different clumps that correspond to an earlier, embedded evolutionary stage of star-forming regions.

\subsubsection{Where do the stars form in $z  \sim 1$ SFGs?}
\label{contrib.sec}

While the analysis in Section\ \ref{clump_nature.sec} sheds light on how star formation activity (SSFR or EW) varies depending on the location within galaxies, it does not reveal how much of the star formation rate integrated over all galaxies is contributed by the center, clump, and disk regimes respectively.  Figure\ \ref{clump_budget.fig} serves to address this question.  We first consider the integrated H$\alpha$ luminosity of all galaxies in the sample, irrespective of whether they have spatial bins in the clump regime or not.  With 8.3\%/79.1\%/12.6\% of the integrated H$\alpha$ luminosity budget being contributed by the central/disk/clump regions (as defined by Eq.\ \ref{center.eq}-\ref{clump.eq}), it is clear that the vast majority of the H$\alpha$ emission arises from the disk component.\footnote{Here, we accounted for the masked pixels that could potentially be affected by [SII] line emission (see Section\ \ref{Halpha_extraction.sec}), to avoid a biased weighting of different radial bins.}  A similar conclusion can be drawn from the breakdown of the rest-frame 2800\AA\ luminosity budget, with modest differences that can be understood from the trends described in Section\ \ref{clump_nature.sec} and specifically Figure\ \ref{HaUV_prof.fig}.  The clump pixels account for 18\% of the integrated $L_{2800}$ luminosity.  It should be noted that the precise percentages quoted above depend significantly on the adopted dividing lines between the regions (Eq.\ \ref{center.eq}-\ref{clump.eq}), and the fractional contribution of excess surface brightness regions should be regarded as coming from bright clumps and spiral features that show sufficient contrast with respect to the extended smooth disk to be detected at the kiloparsec resolution of WFC3.  For example, if we were to lower the dividing line between the disk and clump regions by 0.2 dex, bringing it close to the peak of the contours in Figure 10, the fractional contribution of clumps in this adjusted definition would be 30\% to the H$\alpha$ luminosity and 23\% to the H$\alpha$ derived SFR.  The latter percentages can be interpreted as a rough estimate of the upper limit on the fractional contribution by UV bright clumps.

As discussed extensively in Section\ \ref{calibrating.sec}, the H$\alpha$ luminosity is not a direct tracer of the total on-going star formation, since part of the ionized gas emission is filtered out by the intervening obscuring material.  As discussed by Wuyts et al. (2012), SFGs feature radial gradients in dust extinction.  This results in an enhanced contribution of the inner ($\log(R/R_e) < -0.5$) regions of the galaxies to the integrated SFR budget of our sample: 12.9\% when computing the budget based on dust corrected H$\alpha$ emission.  A similar contribution of 14.8\% is obtained from broad-band SED modeling.  Wuyts et al. (2012) found similar fractional contributions for their mass-complete sample of $z \sim 1$ SFGs, confirming that our H$\alpha$ sample is not subject to significant biases in this regard.  Using broad-band HST imaging, Wuyts et al. (2012) also extended the analysis to $z = 2.5$, finding that the fractional contribution of clumps to the integrated SFR of SFGs increases to $\sim 20\%$ at $z \sim 2$.

We conclude that most stars being formed during and since the peak of cosmic star formation are born in a disk component.  This result is consistent with the analysis of SFR profiles presented for a smaller 3D-HST sample by Nelson et al. (2013), and is in line with a larger body of work based on HST imaging (e.g., F\"orster Schreiber et al. 2011a,b; Wuyts et al. 2011b, 2012; Guo et al. 2012; A. van der Wel et al. in prep) and ground-based kinematic measurements (e.g., Genzel et al. 2006; F\"orster Schreiber et al. 2009, 2013; Lemoine-Busserolle \& Lamareille 2010; Wisnioski et al. 2011; Epinat et al. 2012).

\section {Summary}
\label{summary.sec}

We combined CANDELS high-resolution multi-wavelength imaging on kiloparsec scales of a sample of 473 SFGs at $0.7 < z < 1.5$ with maps of the H$\alpha$ line emission from 3D-HST.  Together, the CANDELS + 3D-HST data shed light on where within galaxies new stars are being formed, and in which amounts stellar mass has been assembled at each location throughout their history prior to observation.  As such, the present study builds on earlier work from 3D-HST (Nelson et al. 2012, 2013) and CANDELS (Wuyts et al. 2012).  Compared to the former study, we expand the sample to a more representative subset of all massive ($\log(M_*)>10$) SFGs at $z \sim 1$, while compared to the latter, we extract sources from all five CANDELS/3D-HST fields instead of one.

In order to optimally exploit the additional constraints from H$\alpha$ on the resolved star formation activity, we started by calibrating empirically the dust corrections to the ionized gas emission required to match the integrated reference SFRs (Wuyts et al. 2011b; for most of our sample UV+PACS or UV+MIPS 24$\mu$m, with values from $U$-to-8 $\mu$m SED modeling for cases without IR detection).  Our empirical calibration (Equation\ \ref{bestfit_dustcor.eq}) highlights the need for an extra extinction correction to H$\alpha$ compared to the underlying continuum light emitted at the same wavelength, although the correction is less severe than proposed by Calzetti et al. (2000) based on nearby starburst galaxies.  Our findings fit in the context of a two-component geometrical dust model, in which the ionized gas emission surrounding young star-forming regions is obscured by dust from the birth cloud itself in addition to the column of diffuse material spread throughout the galaxy that also affects the bulk of galaxy light emitted by older stars (see also Charlot \& Fall 2000; Wild et al. 2011; Pacifici et al. 2012; Chevallard et al. 2013).  As an independent test, we confirmed that this prescription for extra extinction towards HII regions reproduces the relation between the observed H$\alpha$/UV luminosity ratio and the effective visual extinction inferred from broad-band SED modeling.  The fact that the observed H$\alpha$/UV luminosity ratio rises with increasing visual extinction directly implies that H$\alpha$ is a less dust-sensitive SFR tracer than UV emission.

From our resolved stellar population analysis, we draw the following conclusions:

$\bullet$ The H$\alpha$ morphologies of $z \sim 1$ SFGs resemble more closely the ACS $I$-band morphologies than those observed in the WFC3 $H$ band, particularly in the case of large galaxies.

$\bullet$ The H$\alpha$ and rest-frame 2800\AA\ luminosities correlate on a pixel-by-pixel basis, and after proper dust corrections are applied, the SFR estimates based thereupon are internally consistent.  The H$\alpha$/UV luminosity ratios of individual spatial bins also relate to the visual extinction inferred from multi-band HST photometry in a manner that is consistent with the applied correction for extra extinction towards the HII regions.

$\bullet$ We find evidence for the existence of a 'resolved main sequence of star formation': the rate of ongoing star formation per unit area tracks the amount of stellar mass assembled over the same area.  Its near-linear slope is consistent with the one measured for the 'galaxy-integrated main sequence', including an apparent flattening at the high-mass density end, associated with the low-EW inner regions of the most massive galaxies.

$\bullet$ Off-center clumps (or spiral features that show up similarly as regions with excess surface brightness) are characterized by enhanced H$\alpha$ equivalent widths, bluer broad-band colors and correspondingly higher specific SFRs than the underlying disk, implying they are a star formation phenomenon.  Physically, they may correspond to regions with elevated gas fractions and/or star formation efficiencies (Tacconi et al. 2013), where part of the global gas disk collapsed through gravitational instabilities (see, e.g., Bournaud et al. 2008; Genzel et al. 2008; Dekel et al. 2009), although at $z \sim 1$ secular processes on longer timescales, such as spiral density waves, are also likely to contribute.  In an integrated sense, however, the contribution of excess surface brightness regions to the total amount of star formation taking place within $z \sim 1$ SFGs is limited to 10-15\%, depending on the tracer.  Most stars being formed between $z = 1.5$ and $z = 0.7$ are formed in the disk component (see also F\"orster Schreiber et al. 2009, 2013; Nelson et al. 2013).  Wuyts et al. (2012) demonstrated this conclusion holds out to $z \sim 2$, where the fraction of total star formation in off-center clumps increases, but does not exceed $\sim 20\%$.

$\bullet$ Aside from spatial fluctuations in the star formation activity, an inhomogeneous dust distribution can lead to spatial mass-to-light ratio variations, and consequently a range in surface brightness levels at a given radius, even for smooth stellar mass distributions.  The visual extinction gradient inferred from broad-band SED modeling is to first order radial.  We furthermore find clumps/spiral features selected from ACS imaging to exhibit lower H$\alpha$/UV luminosity ratios than the underlying disk.  Interpreting the H$\alpha$/UV ratio as a tracer of extinction, this implies that the clumps/spiral features selected in our study are not just regions of increased star formation activity, but preferentially those observed through a smaller column of obscuring material.

$ $\\
In the past decade, deep HST imaging has transformed our view of high-redshift galaxies from a zoo of faint blobs of light into resolved species whose structure and stellar population content exhibit strong correlations throughout cosmic time (e.g., Franx et al. 2008; Williams et al. 2009; Wuyts et al. 2011b).  More recently, multi-band HST imaging has enabled us to translate the distribution of light within galaxies to more physically relevant distributions of, e.g., stellar mass or star formation (e.g., Wuyts et al. 2012; Guo et al. 2012; Szomoru et al. 2013).  In this paper, we demonstrated that a more complete census of the distribution of baryons within distant galaxies, including the ionized gas as traced by H$\alpha$ (see also Nelson et al. 2012, 2013), helps to constrain spatial variations in star formation activity and dust extinction.  Despite the significant strides forward, especially through the synergy between the CANDELS and 3D-HST legacy programs, a significant fraction of the total baryonic mass budget remains missing from these resolved studies.

In a galaxy-integrated sense, the PHIBSS survey carried out with the Plateau de Bure Interferometer (PdBI) revealed that cold molecular gas traced by CO accounts for a third of the baryonic mass within $z \sim 1.2$ SFGs, with molecular gas mass fractions increasing further out to higher redshifts (Tacconi et al. 2010, 2013; see also Daddi et al. 2010).  Knowing how the molecular gas is distributed within galaxies is crucial to interpret the continuing growth of structure: by quantifying the star formation efficiency in the central, outer disk, and clump/spiral arm regimes separately, identifying the regions with enhanced columns of obscuring material, and finally contrasting the total (gas + stellar) baryonic mass profiles to the available constraints on enclosed mass from kinematics.  

Exploiting the first resolved data set of a distant galaxy consisting of 3D CO and H$\alpha$ data cubes as well as multi-band HST imaging, Genzel et al. (2013) demonstrates the power of such a synergy.  These authors find a correlation between the column of CO line emission and the visual extinction inferred from the HST colors, a near-linear resolved Kennicutt-Schmidt relation, and consistent kinematics of the ionized and molecular gas components.  The superior sensitivity provided by ALMA and the NOEMA extension to PdBI promises to advance this field even further, through mapping systematically the spatial variations in CO line emission as well as the dust-sensitive mm continuum for galaxies spanning a wide dynamic range along and across the main sequence of star formation.

$ $\\

S. W. acknowledges fruitful interactions with Eric F. Bell, Mauro Giavalisco, and Benjamin J. Weiner.  We also thank the anonymous referee for his/her constructive comments.
Support for Program number HST-GO-12060 and HST-GO-12177 was provided by NASA through a grant from the Space Telescope Science Institute, which is operated by the Association of Universities for Research in Astronomy, Incorporated, under NASA contract NAS5-26555.

$ $\\

\onecolumngrid

\vspace{0.2in}
\begin{center} APPENDIX  Robustness of the resolved SED modeling\end{center}

As described in Section\ \ref{CANDELS.sec}, the available high-resolution HST imagery in the CANDELS/3D-HST fields varies from 7-band $BVizYJH$ photometry in the GOODS fields, contributing 46\% of our sample, to a more modest sampling of the SED, in $VIJH$, for EGS, UDS and COSMOS (objects with less coverage were excluded from the sample).  Our choice to include also the latter three fields in the analysis is therefore necessarily a trade-off between having the richest possible set of information per object, and maximizing the sample statistics.

We verify the validity of our approach in two ways.  First, we repeat our entire analysis five times, each time using the data set of an individual field separately.  We revisit the relation between H$\alpha$- and broad-band derived properties (Section\ \ref{broad_vs_line.sec}), the variation of stellar population properties across the co-added rest-UV surface brightness profiles (Section\ \ref{clump_nature.sec}), as well as the fractional contributions to the H$\alpha$ luminosity and integrated SFR by the central/disk/clump regions (Section\ \ref{contrib.sec}).  Doing so, we confirm that the same generic behavior is seen in each of the fields separately, albeit naturally based on smaller number statistics.  This test boosts our confidence that the conclusions drawn in this paper are not driven by just a limited subset of the objects entering our analysis.  For example, the relative fractional contributions of galaxy centers, disks, and clumps or spiral arms, vary by at most a few percent between fields.

Second, we tested the impact of having 4 rather than 7 flux points per spatial bin, by repeating the analysis for the GOODS-South field using only the $ViJH$ resolved photometry.  We found that the best convergence between solutions of 4- and 7-band fitting is obtained when assigning a weight $w_i = e^{-\chi_i^2/2}$ to each grid cell and marginalizing over the multi-dimensional parameter space explored in the fitting process (see Salim et al. 2007 for an extensive discussion).  We therefore chose to follow this approach throughout the paper, but note that consistent conclusions are reached when using the minimum $\chi^2$ value.  Comparing the 4- to 7-band fitting, we reproduce the overall lessons learned on resolved star formation patterns at $z \sim 1$ presented in this paper: the existence of a resolved main sequence of star formation with similar slope as the galaxy-integrated version, and excursions from this relation in the form of off-center regions with excess star formation, or low equivalent width regions in galaxy centers.  Quantitatively, the stellar surface mass density, rest-frame 2800\AA\ surface brightness, and rest-frame $(U-V)_{rest}$ color are least affected by the transition from 7 to 4 bands, inducing less than 0.07 dex scatter and no systematic offsets.  The absence of systematics is most crucial, as throughout the paper we never treat individual spatial bins separately, but instead always consider statistics averaged over large numbers of spatial bins taken from all objects in the sample.

Unsurprisingly, the inferred age, SFR, and visual extinction are less well constrained by 4-band SEDs, with scatter in the 4- to 7-band comparison of 0.18 dex, 0.20 dex and 0.32 mag respectively, and strong (anti-)correlations between the offsets.  However, also for these parameters we measure only minor systematic offsets, with amplitudes that are a factor 6 smaller than the observed scatter.  This propagates to variations in the co-added surface brightness profiles and resolved main sequence of comparable or even smaller amplitude than the variations seen from field to field.

We conclude that, to the level of detail presented in this paper, the above tests support the robustness of our results.  However, in order to push the study of resolved stellar populations at high redshift further, toward extensive analyses on a per object basis, additional means to better disentangle age and dust effects would be desirable.  Two possible avenues are the inclusion of WFC3 UVIS data (Teplitz et al. 2013), providing access to resolved maps of the UV slope $\beta$, and the upcoming JWST/NIRCam mission, which will enable the application of the UVJ diagnostic technique (Wuyts et al. 2007) on subgalactic scales.

$ $\\
\begin{references}
{\footnotesize

\reference{} Abraham, R. G., Ellis, R. S., Fabian, A. C., Tanvir, N.,\& Glazebrook, K. 1999, MNRAS, 303, 641
\reference{} Barro, G., Faber, S. M., P\'erez-Gonz\'alez, P. G., et al. 2013, ApJ, 765, 104
\reference{} Bell, E. F.,\& de Jong, R. S. 2001, ApJ, 550, 212
\reference{} Bell, E. F., van der Wel, A., Papovich, C., et al. 2012, ApJ, 753, 167
\reference{} Bournaud, F., Daddi, E., Elmegreen, B. G., et al. 2008, A\&A, 486, 741
\reference{} Bournaud, F., Dekel, A., Teyssier, R., Cacciato, M., Daddi, E., Juneau, S.,\& Shankar, F. 2011, ApJ, 741, 33
\reference{} Brammer, G. B., van Dokkum, P. G., Franx, M., et al. 2012, ApJS, 200, 13
\reference{} Bruce, V. A., Dunlop, J. S., Cirasuolo, M., et al. 2012, MNRAS, 427, 1666
\reference{} Bruzual, G,\& Charlot, S. 2003, MNRAS, 344, 1000
\reference{} Cacciato, M., Dekel, A.,\& Genel, S. 2012, MNRAS, 421, 818
\reference{} Calzetti, D., Kenicutt, R. C., Bianchi, L., et al. 2005, ApJ, 633, 871
\reference{} Calzetti, D., Armus, L., Bohlin, R. C., Kinney, A. L., Koornneef, J.\& Storchi-Bergmann, T. 2000, ApJ, 533, 682
\reference{} Calzetti, D., Kinney, A. L.,\& Storchi-Bergmann, T. 1994, ApJ, 429, 582
\reference{} Cappellari, M.,\& Copin, Y. 2003, MNRAS, 342, 345
\reference{} Chabrier, G. 2003, PASP, 115, 763
\reference{} Charlot, S.,\& Fall, S. M. 2000, ApJ, 539, 718
\reference{} Chevallard, J., Charlot, S., Wandelt, B.,\& Wild, V. 2013, MNRAS, in press (arXiv1303.6631)
\reference{} Daddi, E., et al. 2007, ApJ, 670, 156
\reference{} Daddi, E., Bournaud, F., Walter, F., et al. 2010, ApJ, 713, 686
\reference{} Daddi, E., Cimatti, A., Renzini, A., Fontana, A., Mignoli, M., Pozzetti, L., Tozzi, P.,\& Zamorani, G. 2004, ApJ, 617, 746
\reference{} Davis, M., Guhathakurta, P., Konidaris, N. P., et al. 2007, ApJ, 660, 1
\reference{} Dekel, A., Sari, R.,\& Ceverino, D. 2009, ApJ, 703, 785
\reference{} Dickinson, M., Hanley, C., Elston, R., et al. 2000, ApJ, 531, 624
\reference{} Elbaz, D., et al. 2007, A\&A, 468, 33
\reference{} Elbaz, D., et al. 2011, A\&A, 533, 119
\reference{} Elmegreen, B. G., Elmegreen, D. M., Fernandez, M. X.,\& Lemonias, J. J. 2009b, ApJ, 692, 12
\reference{} Elmegreen, D. M., Elmegreen, B. G., Marcus, M. T., Shahinyan, K., Yau, A.,\& Peterson, M. 2009a, ApJ, 701, 306
\reference{} Elmegreen, D. M., Elmegreen, B. G.,\& Hirst, A. C. 2004, ApJ, 604, 21
\reference{} Epinat, B., Tasca, L., Amram, P., et al. 2012, A\&A, 539, 92
\reference{} F\"{o}rster Schreiber, N. M., Shapley, A. E., Erb, D. K., ApJ, 731, 65
\reference{} F\"{o}rster Schreiber, N. M., Shapley, A. E., Genzel, R., et al. 2011b, ApJ, 739, 45
\reference{} F\"{o}rster Schreiber, N. M., Genzel, R., Bouch\'{e}, N., et al. 2009, ApJ, 706, 1364
\reference{} F\"orster Schreiber, N. M., van Dokkum, P. G., Franx, M., et al. 2004, ApJ, 616, 40
\reference{} Franx, M., van Dokkum, P. G., F\"orster Schreiber, N. M., Wuyts, S., Labb\'e, I.,\& Toft, S. 2008, ApJ, 688, 770
\reference{} Franx, M., Labb\'{e}, I., Rudnick, G., et al. 2003, ApJ, 587, 79
\reference{} Genel, S., Naab, T., Genzel, R., et al. 2012, ApJ, 745, 11
\reference{} Genzel, R., Tacconi, L. J., Kurk, J., et al. 2013, submitted to ApJ (arXiv1304.0668)
\reference{} Genzel, R., Newman, S., Jones, T., et al. 2011, ApJ, 733, 101
\reference{} Genzel, R., Burkert, A., Bouch\'{e}, N., et al. 2008, ApJ, 687, 59
\reference{} Genzel, R., Tacconi, L. J., Eisenhauer, F., et al. 2006, Nature, 442, 786
\reference{} Giavalisco, M., Ferguson, H. C., Koekemoer, A. M., et al. 2004, ApJ, 600, 93
\reference{} Grogin, N. A., Kocevski, D. D., Faber, S. M., et al. 2011, ApJS, 197, 35
\reference{} Guo, Y., Giavalisco, M., Ferguson, H. C., Cassata, P.,\& Koekemoer, A. M. 2012, ApJ, 757, 120
\reference{} Hopkins, P. F., Keres, D., Murray, N., Quataert, E.,\& Hernquist, L. 2012, MNRAS, 427, 968
\reference{} Hummer, D. G.,\& Storey, P. J. 1987, MNRAS, 224, 801
\reference{} Ilbert, O., Capak, P., Salvato, M., et al. 2009, ApJ, 690, 1236
\reference{} Immeli, A., Samland, M., Gerhard, O.,\& Westera, P. 2004a, A\&A, 413, 547
\reference{} Immeli, A., Samland, M., Westera, P.,\& Gerhard, O. 2004b, ApJ, 611, 20
\reference{} Jones, T., Ellis, R. S., Richard, J.,\& Jullo, E. 2013, ApJ, 765, 48
\reference{} Kartaltepe, J. S., Dickinson, M. Alexander, D. M., et al. 2012, ApJ, 757, 23
\reference{} Kennicutt, R. C. 1998, ARA\&A, 36, 189
\reference{} Koekemoer, A. M., Faber, S. M., Ferguson, H. C., 2011, ApJS, 197, 36
\reference{} Koekemoer, A. M., Aussel, H., Calzetti, D., et al. 2007, ApJS, 172, 196
\reference{} Kriek, M., van Dokkum, P. G., Franx, M., Illingworth, G. D.,\& Magee, D. K. 2009a, ApJ, 705, 71
\reference{} Kriek, M., van Dokkum, P. G., Labb\'{e}, I., Franx, M., Illingworth, G. D., Marchesini, D.,\& Quadri, R. F. 2009b, ApJ, 700, 221
\reference{} Krumholz, M,\& Burkert, A. 2010, ApJ, 724, 895
\reference{} Lanyon-Foster, M. M., Conselice, C. J.,\& Merrifield, M. R. 2012, MNRAS, 424, 1852
\reference{} Larson, R. B. 1981, MNRAS, 194, 809
\reference{} Law, D. R., Shapley, A. E., Steidel, C. C., Reddy, N. A., Christensen, C. R.,\& Erb, D. K. 2012, Nature, 487, 338
\reference{} Lee, S.-K., Idzi, R., Ferguson, H. C., Somerville, R. S., Wiklind, T.,\& Giavalisco, M. 2009, ApJS, 184, 100
\reference{} Lemoine-Busserolle, M.,\& Lamareille, F. 2010, MNRAS, 402, 2291
\reference{} Liu, X., Shapley, A. E., Coil, A. L., Brinchmann, J.,\& Ma, C.-P. 2008, ApJ, 678, 758
\reference{} Ly, C., Malkan, M. A., Kashikawa, N., Ota, K., Shimasaku, K., Iye, M.,\& Currie, T. 2012, ApJ, 747, 16
\reference{} Magnelli, B., Popesso, P., Berta, S., et al. 2013, A\&A, in press (arXiv1303.4436)
\reference{} Mancini, C., F\"orster Schreiber, N. M., Renzini, A., et al. 2011, ApJ, 743, 86
\reference{} Meurer, G. R., Wong, O. I., Kim, J. H., et al. 2009, ApJ, 695, 765
\reference{} Mitchell, P. D., Lacey, C. G., Baugh, C. M.,\& Cole, S. 2013, submitted to MNRAS (arXiv1303.7228)
\reference{} Nelson, E. J., van Dokkum, P. G., Momcheva, I., et al. 2013, ApJ, 763, 16
\reference{} Nelson, E. J., van Dokkum, P. G., Brammer, G., et al. 2012, ApJ, 747, 28
\reference{} Newman, S. F., Shapiro Griffin, K., Genzel, R., et al. 2012, ApJ, 752, 111
\reference{} Noeske, K. G., et al. 2007, ApJ, 660, 43
\reference{} Noguchi, M. 1999, ApJ, 514, 77
\reference{} Pacifici, C., Charlot, S., Blaizot, J.,\& Brinchmann, J. 2012, MNRAS, 421, 2002
\reference{} Papovich, C., Dickinson, M.,\& Ferguson, H. C. 2001, ApJ, 559, 620
\reference{} Pforr, J., Maraston, C.,\& Tonini, C. 2012, MNRAS, 422, 3285
\reference{} Queyrel, J., Contini, T., Kissler-Patig, M., et al. 2012, A\&A, 539, 93
\reference{} Rix, H.-W., Barden, M., Beckwith, S. V. W., et al. 2004, ApJS, 152, 163
\reference{} Salim, S., Rich, R. M., Charlot, S., et al. 2007, ApJS, 173, 267
\reference{} Santini, P., Rosario, D. J., Shao, L., et al. 2012, A\&A, 540, 109
\reference{} Shapley, A. E., Steidel, C. C., Erb, D. K., Reddy, N. A., Adelberger, K. L., Pettini, M., Barmby, P.,\& Huang, J. 2005, ApJ, 626, 698
\reference{} Spitler, L. R., Labb\'e, I., Glazebrook, K., et al. 2012, ApJ, 748, 21
\reference{} Steidel, C. C., Giavalisco, M., Pettini, M., Dickinson, M.,\& Adelberger, K. L. 1996, ApJ, 462, 17
\reference{} Swinbank, A. M., Sobral, D., Smail, I., Geach, J. E., Best, P. N., McCarthy, I. G., Crain, R. A.,\& Theuns, T. 2012, MNRAS, 426, 935
\reference{} Szomoru, D., Franx, M., van Dokkum, P. G., Trenti, M., Illingworth, G. D., Labb\'e, I.,\& Oesch, P. 2013, ApJ, 763, 73
\reference{} Szomoru, D., Franx, M., Bouwens, R. J., van Dokkum, P. G., Labb\'e, I., Illingworth, G. D.,\& Trenti, M. 2011, ApJ, 735, 22
\reference{} Tacconi, L. J., Neri, R., Genzel, R., et al. 2013, ApJ, 768, 74
\reference{} Tacconi, L. J., Genzel, R., Neri, R., et al. 2010, Nature, 463, 781
\reference{} Taylor, E. N., Hopkins, A. M., Baldry, I. K., et al. 2011, MNRAS, 418, 1587
\reference{} Teplitz, H. I., Rafelski, M., Kurczynski, P., et al. 2013, AJ, submitted (arXiv1305.1357)
\reference{} Thompson, R. I., Storrie-Lombardi, L. J., Weymann, R. J., et al. 1999, AJ, 117, 17
\reference{} Trump, J. R., Konidaris, N. P., Barro, G., et al. 2013, ApJ, 763, 6
\reference{} van der Wel, A., Bell, E. F., H\"{a}ussler, B., et al. 2012, ApJS, 203, 24
\reference{} van Dokkum, P. G., Brammer, G., Fumagalli, M., et al. 2011, ApJ, 743, 15
\reference{} van Dokkum, P. G., Franx, M., F\"orster Schreiber, N. M., et al. 2004, ApJ, 611, 703
\reference{} Wang, T., Huang, J., Faber, S. M., et al. 2012, ApJ, 752, 134
\reference{} Whitaker, K. E., van Dokkum, P. G., Brammer, G.,\& Franx, M. 2012, ApJ, 754, 29
\reference{} Whitaker, K. E., Labb\'e, I., van Dokkum, P. G., et al. 2011, ApJ, 735, 86
\reference{} Wild, V., Charlot, S., Brinchmann, J., et al. 2011, MNRAS, 417, 1760
\reference{} Williams, R. J., Quadri, R. F., Franx, M., van Dokkum, P. G.,\& Labb\'e, I. 2009, ApJ, 691, 1879
\reference{} Williams, R. E., Blacker, B., Dickinson, M., et al. 1996, AJ, 112, 1335
\reference{} Wisnioski, E., Glazebrook, K., Blake, C., et al. 2011, MNRAS, 417, 2601
\reference{} Wuyts, S., F\"{o}rster Schreiber, N. M., Genzel, R., et al. 2012, ApJ, 753, 114
\reference{} Wuyts, S., F\"{o}rster Schreiber, N. M., van der Wel, A., et al. 2011b, ApJ, 742, 96
\reference{} Wuyts, S., F\"{o}rster Schreiber, N. M., Lutz, D., et al. 2011a, ApJ, 738, 106
\reference{} Wuyts, S., Franx, M., Cox, T. J., Hernquist, L., Hopkins, P. F., Robertson, B. E.,\& van Dokkum, P. G. 2009, ApJ, 696, 348
\reference{} Wuyts, S. Labb\'e, I., Franx, M., et al. 2007, ApJ, 655, 51
\reference{} Yuan, T.-T., Kewley, L. J.,\& Richard, J. 2013, ApJ, 763, 9
\reference{} Yuan, T.-T., Kewley, L. J., Swinbank, A. M.,\& Richard, J. 2012, ApJ, 759, 66

}
\end {references}

\end {document}